\documentclass[12pt,letterpaper]{article}
\pdfoutput=1
\usepackage{graphicx}
\usepackage{subfig}
\usepackage{booktabs}
\usepackage{amsfonts,amssymb,amsmath}

\newcommand{\be}{\begin{equation}}
\newcommand{\ee}{\end{equation}}
\newcommand{\ben}{\begin{displaymath}}
\newcommand{\een}{\end{displaymath}}
\newcommand{\bea}{\begin{eqnarray}}
\newcommand{\eea}{\end{eqnarray}}
\newcommand{\bean}{\begin{eqnarray*}}
\newcommand{\eean}{\end{eqnarray*}}

\def\a {\alpha}

\def\s {\sigma}

\newcommand{\eg}{{\it e.g.}}
\newcommand{\ie}{{\it i.e.}}

\newcommand{\tr}{\mbox{Tr}}

\newcommand{\commentout}[1]{}

\numberwithin{equation}{section}

\newcommand{\beq}{\begin{equation}}
\newcommand{\eeq}{\end{equation}}
\newcommand{\beqr}{\begin{displaymath}}
\newcommand{\eeqr}{\end{displaymath}}
\newcommand{\beqa}{\begin{eqnarray}}
\newcommand{\eeqa}{\end{eqnarray}}
\newcommand{\beqar}{\begin{eqnarray*}}
\newcommand{\eeqar}{\end{eqnarray*}}
\newcommand{\cN}{{\cal N}}
\newcommand{\cD}{{\cal D}}
\newcommand{\cO}{{\cal O}}

\newcommand{\non}{\nonumber}

\newcommand{\cA}{{\cal A}}

\newcommand{\cC}{{\cal C}}
\newcommand{\cW}{{\cal W}}

\newcommand{\half}{\ensuremath{\frac{1}{2}}}

\newcommand{\ba}{\ensuremath{\bar{a}}}

\newcommand{\N}[1]{\ensuremath{\cN=#1}}

\def\be{\begin{equation}}
\def\ee{\end{equation}}
\def\ba{\begin{eqnarray}}
\def\ea{\end{eqnarray}}

\def\Xint#1{\mathchoice
{\XXint\displaystyle\textstyle{#1}}%
{\XXint\textstyle\scriptstyle{#1}}%
{\XXint\scriptstyle\scriptscriptstyle{#1}}%
{\XXint\scriptscriptstyle\scriptscriptstyle{#1}}%
\!\int}
\def\XXint#1#2#3{{\setbox0=\hbox{$#1{#2#3}{\int}$ }
\vcenter{\hbox{$#2#3$ }}\kern-.6\wd0}}

\def\dashint{\Xint-}

\newcommand{\rhoM}{\hat{\rho}}

\newcommand{\bmu}{\ensuremath{\bar{\mu}}}
\newcommand{\bnu}{\ensuremath{\bar{\nu}}}

\newcommand{\re}{\ensuremath{\mathrm{Re}}}

\DeclareMathOperator{\Tr}{Tr}

\newcommand{\Dbm}[1]{\ensuremath{\bar{\mathcal{D}}^{(-)}_{#1}}}
\newcommand{\U}[1]{\ensuremath{\mathcal{U}_{#1}}}

\newcommand{\Ub}[1]{\ensuremath{\bar{\mathcal{U}}_{#1}}}

\newcommand{\F}[1]{\ensuremath{\mathcal{F}_{#1}}}

\newcommand{\Nfour}{$\mathcal{N} = 4$ SYM }

\begin{document}


\title{\LARGE \bf Loop Equations and bootstrap methods in the lattice}

\author{Peter D. Anderson$^{1,2}$ and Martin Kruczenski$^1$
\thanks{E-mail: \texttt{ander324@purdue.edu, markru@purdue.edu}}\\
        $^1$ Department of Physics and Astronomy, Purdue U.,  \\
        525 Northwestern Avenue, W. Lafayette, IN 47907-2036,  USA\\
        $^2$ Wigner Research Center for Physics of the HAS, \\
        29--33 Konkoly--Thege Mikl\'os Str. H-1121 Budapest, Hungary }

\maketitle

\vspace{-1cm}

\begin{abstract}
 Pure gauge theories can be formulated in terms of Wilson Loops correlators by means of the loop equation. In the large-N limit this equation closes in the expectation value of single loops. In particular, using the lattice as a regulator, it becomes a well defined equation for a discrete set of loops. In this paper we study different numerical approaches to solving this equation. Previous ideas gave good results in the strong coupling region. Here we propose an alternative method based on the observation that certain matrices $\rhoM$ of  Wilson loop expectation values are positive definite. They also have unit trace ($\rhoM\succeq0, \tr\rhoM=1$), in fact they can be defined as density matrices in the space of open loops after tracing over color indices and can be used to define an entropy associated with the loss of information due to such trace $S_{WL}=-\tr[ \rhoM\ln \rhoM]$. 
 The condition that such matrices are positive definite allows us to study the weak coupling region which is relevant for the continuum limit. In the exactly solvable case of two dimensions this approach gives 
very good results by considering just a few loops. In four dimensions it gives good results in the weak coupling region and therefore is complementary to the strong coupling expansion. We compare the results with standard Monte Carlo simulations. 
\end{abstract}

\clearpage
\newpage



\section{Introduction}
\label{intro}
 
   Gauge theories are of fundamental importance for our understanding of Nature but many of their properties are still mysterious, for example in pure gauge theories, the phenomenon of
confinement is still not fully understood. Even further,  the AdS/CFT correspondence \cite{malda} has shown that gauge theories in the strongly coupled regime can be  described equally well 
in terms of string theory in a higher dimensional space. That means that certain gauge theories contain quantum gravity, emergent space-time and strings as bound states. Fundamental to this understanding is the relation of gauge theories and string theory in the limit of a large number of colors as envisioned by 't Hooft \cite{largeN}. In such relation, exemplified by AdS/CFT, the string theory description is completely in terms of gauge invariant operators. In fact the gauge symmetry is not a symmetry of the dual theory in accordance with the usual understanding that a gauge symmetry is a manifestation of redundant degrees of freedoms that have to be eliminated. Taken to its logical conclusion, the principle of gauge invariance cannot be used as the basis to construct such a theory and perhaps more ideas are needed to understand the fundamental principles lying behind gauge theories.  
 
  For these reasons it is natural to study gauge invariant formulations of gauge theories. In fact it is known that gauge theories can be formulated entirely in terms of Wilson loops. In particular, in the large-N limit Wilson loops obey an equation that closes in the expectation value of single Wilson loops. This is known as the loop equation \cite{loopeqn,review1984} or the Migdal--Makeenko equation. For finite N the equation is also valid but closes in the expectation value of disconnected (\ie\ multitrace) Wilson loops. Since it is not clear how to renormalize the loop equation it can be better study perturbatively or in the lattice. Motivated by this, here we discuss different numerical and analytical methods to study the loop equation in the lattice. 

 Although, as we argued, this equation is of great importance there does not seem to be many studies on how to solve it. A notable exception is the very interesting work by Marchesini \cite{Marchesini} where a formal solution and a numerical approach to solve the loop equation was proposed. We discuss this approach in detail but unfortunately it seems restricted to the strongly coupled regime, moreover, and the known numerical results are for the 2d case. The method we propose is based on the observation that certain matrices constructed of Wilson loop expectation values are positive definite. They can be thought of as reduced density matrices in the space of open loops after tracing over color indices. Imposing this extra condition allow us to select valid solutions of the loop equations. We call this approach a bootstrap approach since it uses general positivity properties of the theory to impose bounds on the solutions and also since imposing positive definiteness is an important part of the recently developed and highly successful conformal bootstrap program \cite{CFTbootstrap,sdpaCFT}\footnote{We want to clarify however, that the ideas discussed here do not seem to have any relation with conformal symmetry. Perhaps closer is the idea of applying the bootstrap method to non-conformal theories \cite{nonCFTbootstrap} but we do not know of any direct relation with the present work.}. The bounds we impose are for the expectation value of the energy. In two dimensions such bounds constrain the solution to be equal to the exact solution with high degree of accuracy. 
 In four dimensions the bounds are less restrictive (due to larger computational complexity) but we can resort to a simple approximation, at small coupling we minimize the expectation value of the energy subject to the constraints, whereas at large coupling the entropy should be maximized. This gives results which are in good agreement with simulations and a reasonable approximation to the coupling where the transition occurs. The weak coupling region is well described by this method although, at the moment, this is not enough to understand the continuum limit. 

  For future work, it seems of great interest to apply this method to \N{4} SYM in order to make contact with the AdS/CFT correspondence. In this paper we take a few initial steps in this direction by briefly considering the bosonic sector of \N{4} SYM but leave a detailed study for future work.
 
 It is interesting to note that the relevance of the extra positivity conditions at weak coupling was already observed \cite{Jevicki} in the collective field method of Jevicki and Sakita \cite{CollectiveField}. Using the Kogut-Susskind approach a Lattice Hamiltonian in loop space was derived and numerically studied in \cite{Jevicki}. See also the related work by Yaffe \cite{Yaffe} using coherent states. 
 
 This paper is organized as follows, in the next section we review the derivation of the loop equation and summarize the main ideas presented in this paper, following that we consider in great detail the two dimensional case since its solution is known exactly and can be used to test various approaches very easily. Afterwards, we apply those ideas to the four dimensional case and show that our numerical approach gives a good understanding of the gauge theory in the small coupling regime relevant for the continuum limit. Finally we describe a numerical simulation used to validate the results,
 discuss briefly the case of \N{4} SYM and conclude with a summary of the results and possible extensions and improvements.

\section{Lattice gauge theory, a brief summary.}
\label{lgt}

 In this section we consider a four dimensional $SU(N)$ gauge theory in an infinite cubic lattice with Wilson action and briefly review known results for the large N limit. Then we discuss the derivation of the loop equation and introduce the notation we use in this paper. There is an extensive literature on the subject, our presentation here is just to summarize known results that are needed later in the paper and mostly follow the classic review \cite{Drouffe} and the book \cite{MakeenkoBook} as regards to the loop equation.

\subsection{Lattice action and known results}\label{lgt1}

 The system we consider is a cubic lattice where to each oriented link is associated a matrix  $U_\mu \in SU(N)$. To the same link with opposite orientation we associate the matrix $U_{\bmu}=U_\mu^\dagger$. The action is the Wilson action
\beq
 S = -\frac{N}{2\lambda} \sum_{P} \tr U_P \ ,
 \label{a1}
\eeq
 where the sum is over all oriented plaquettes $P$. Here $U_P$ is the product of the four matrices associated with the plaquette and oriented means that we sum the trace of both possible orientation so that the action is real.
 The partition function is
 \beq
 Z = \int \prod_{\vec{x},\mu} dU_\mu(\vec{x})\ e^{-S} .
 \label{a2}
\eeq
 In four dimensions, for $N\ge4$ numerical results indicate that this theory has a first order phase transition as a function of $\lambda$ \cite{Drouffe}. In the large-N limit the transition occurs at $\lambda_c = 1.3904$, as computed using the Twisted Eguchi-Kawai (TEK) model\cite{Campostrini:1998zd}. The nature of the transition is easily understood by considering the partition function in eq.(\ref{a2}) as defining a classical four dimensional statistical system with Hamiltonian 
 \beq
  H = -\frac{N}{2} \sum_{P} \tr U_P \ ,
  \label{a3}
 \eeq
 and temperature $T=\lambda$. At small temperature $\lambda\rightarrow 0$ we minimize the energy, \ie\ the links $U_\mu$ fluctuate around gauge trivial configurations. At large temperature $\lambda\rightarrow \infty$ the entropy should be maximized and the links variables $U_\mu$ explore all possible values with equal probability. Thus, the transition is a typical first order first transition between ordered and disordered states. 
 
 The large coupling phase is confining and easily studied analytically in terms of a strong coupling expansion already proposed by Wilson\cite{Wilson:1974sk}. On the other hand the continuum limit is obtained in the region $\lambda\rightarrow 0$ which is more difficult to study.  
A Wilson loop expectation value is a real number associated with a closed path in the lattice and defined as
\beq
\cW_{\cC} =\frac{1}{Z} \int \prod_{\vec{x}\mu} dU_\mu(\vec{x})\ \frac{1}{N} \tr(U_{\mu_1}\ldots U_{\mu_L})e^{-S} \ ,
\label{a4}
\eeq
where inside the trace we multiplied in cyclic order all the matrices associated to the given path $\cC=\{\mu_1\mu_2\ldots \mu_L\}$.

\subsubsection{Strong coupling phase} \label{lgt2}
Analytically, the Wilson loop expectation values $W_{\cC}$ can be compute in a strong coupling expansion $\lambda\gg 1$ by expanding the exponential of the action. The result has an interesting interpretation in terms of a sum over surfaces ending on the loop \cite{surfaces}. In 4 dimensions and in the large N-limit the expectation value for the plaquette is given by
\beq
 \cW_1 = u = \frac{1}{2\lambda}+\frac{1}{8\lambda^5}+\frac{15}{128}\frac{1}{\lambda^9}+\frac{17}{256}\frac{1}{\lambda^{11}}+\frac{273}{2048}\frac{1}{\lambda^{13}}+\frac{185}{1024}\frac{1}{\lambda^{15}} +\ldots
 \label{a5}
\eeq 
where we called the plaquette as Wilson loop one $\cW_1$ also denoted as $u$. Wilson loop zero is the single point or null loop that obeys $\cW_0=1$. Such large orders in perturbation theory can be computed by using a character expansion of the exponential. Notice that $\cW_1=u$ is of particular importance since it determines the average Energy density (energy per lattice site)
\beq
 \frac{E}{VN^2} = - 6 u \ ,
 \label{a6}
\eeq
where $V$ is the number of sites and the factor six is the number of plaquettes per site computed as $d(d-1)/2$, in  dimension $d=4$.

\subsubsection{Weak coupling} \label{lgt3}
For small coupling the $U_\mu$ matrices only have small fluctuations around the identity (up to gauge transformations). In that case one can write the theory in terms of a hermitian gauge field $A_\mu$, $U_\mu=e^{iA_\mu}$ and use Feynman diagrams to compute Wilson loops. The result for the plaquette in four dimensions and the large N limit is \cite{Heller:1984hx} 
\beq
  W_1 = u = 1-\frac{\lambda}{4} - \left(\frac{1}{48}+ c_2\right)\lambda^2,  \ \ \ \ c_2 \simeq -0.00041 \ .
  \label{a7}
\eeq
 Of course perturbation becomes unreliable for loops of large area since it does not capture the phenomenon of confinement that implies that such loops obey the area law. 
 
\subsubsection{Transition in mean field approximation} \label{lgt4}
 
 The transition from weak to the strong coupling regime is a first order transition that can be understood by a simple mean field approximation \cite{Drouffe}. Within this approximation and in axial gauge, the free energy per site is given by \cite{Drouffe}
 \beq
  a = -\frac{\lambda}{V N^2} \ln Z =  \left\{
  \begin{array}{lcl}
    3\left(1-\frac{1}{\lambda}\right) v^2 - \frac{3}{\lambda} v^4 &\ \  &    0<v<\half \\
    \frac{3}{4}-\frac{3}{2}\ln(2(1-v)) - \frac{3}{\lambda}(v^4+v^2) & &\half <v<1 \ ,
  \end{array}
   \right.
   \label{a8}
 \eeq
where $v$ is a free parameter that is fixed by minimizing $a$, \ie $\frac{\partial a}{\partial v}=0$. The function $a(v)$ has a minimum at $v=0$ and for $\lambda< \lambda_M\simeq 1.68$ has a second minimum corresponding to the small coupling phase that has lower free energy for $\lambda<\lambda_c\simeq 1.48$. The expectation value of the plaquette is given by
\beq
 u =-\frac{1}{6} \left(a-\lambda\frac{\partial a}{\partial \lambda}\right) = \half(v^2+v^4) \ ,
 \label{a9}
\eeq
 as can be derived from the partition function. In the strong coupling regime we get the very crude approximation $u=0$, in fact all Wilson loops vanish. In the small coupling regime we get a non-trivial function that can be expanded around $\lambda=0$ as
 \beq
  u = 1-\frac{1}{4}\lambda +\ldots
  \label{a10}
 \eeq 
This agrees with eq.(\ref{a7}) but the $\lambda^2$ term ($-\frac{7}{288} \lambda^2$ ) is already incorrect. The mean field approximation can be improved \cite{Drouffe} but we just wanted to emphasize that a simple approach captures the important physics. 
 
\subsubsection{Numerical simulations} \label{lgt5}

\begin{figure}
\centering
\includegraphics[width=13cm]{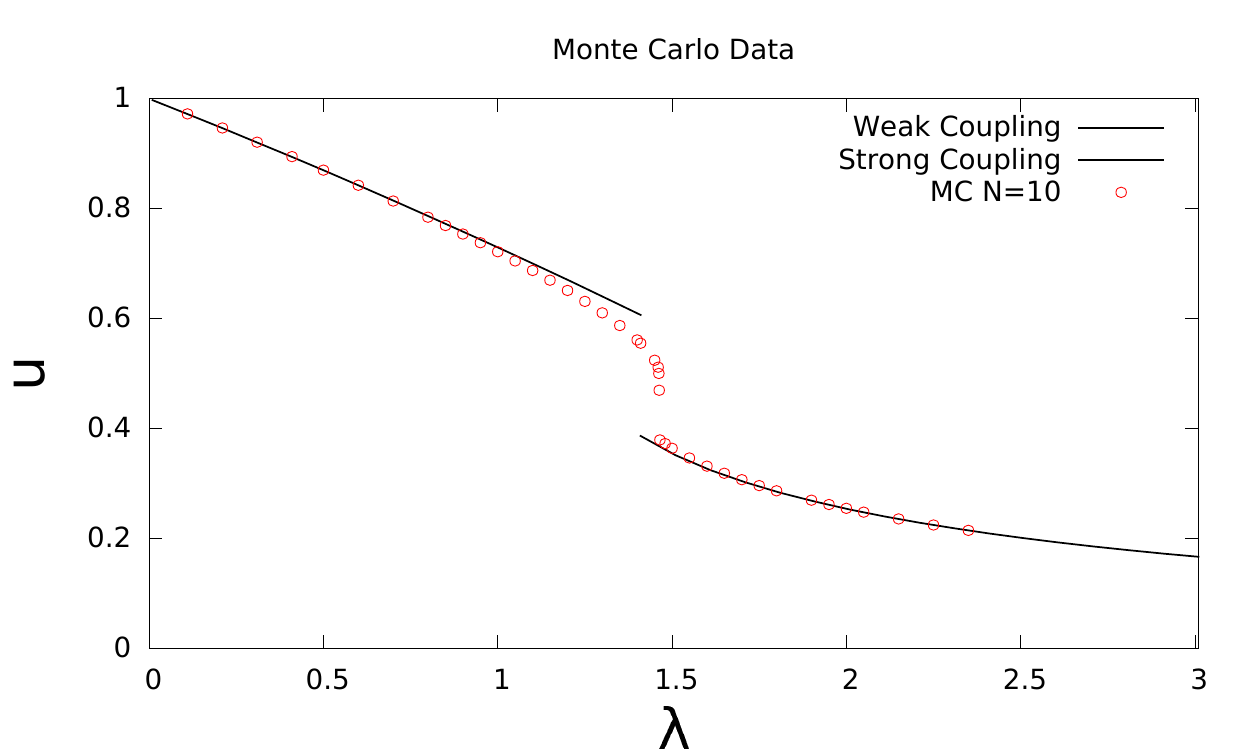}
\caption{Monte Carlo results for the plaquette verse the t' Hooft coupling}
\label{fig:MC}
\end{figure}
 
 All previous results are for infinite lattices and in the strict $N\rightarrow \infty$ limit. For finite N and small lattices one can use numerical simulations and extrapolate the results to infinite N. The first results in this direction were by Creutz and Moriarty \cite{Creutz} and more recently by Meyer and Teper \cite{Teper}. As part of this work we performed a numerical simulation for $N=10$ in an $8^4$ lattice which allowed us to check various ideas regarding the loop equation. The results for the expectation value of the plaquette are displayed in fig.\ref{fig:MC} where a very good agreement is seen with the perturbative and large coupling expansion in their regimes of validity. The position of the phase transition for $SU(10)$ is seen to be around $\lambda_c(N=10)\simeq1.46$ in agreement with the literature and close to the large-N value 
 $\lambda_c \simeq 1.3904$ \cite{Campostrini:1998zd}. 
  
 \subsubsection{Summary} \label{lgt6}
 Clearly there is already a very good and detailed understanding of this system, the strong and weak coupling regimes can be understood by series expansion and the transition using mean field. Numerical simulations validate the whole picture as
 summarized in fig.\ref{fig:MC}. It should be emphasized however that the main physical interest lies in the continuum limit that appears in the $\lambda\rightarrow 0$ region. In practice one has to show that large loops obey the area law in the weak coupling phase, a result that cannot be obtained by perturbation theory and can only be found by extrapolation of numerical simulations. 
  
In any case, our intention here is to study this system purely in terms of gauge invariant operators, namely with no reference to the variables $U_\mu$. For that reason we review now the derivation of the loop equation.
 
\subsection{The loop equation} \label{lgt7}

\begin{figure}
 \centering
 \includegraphics[width=10cm]{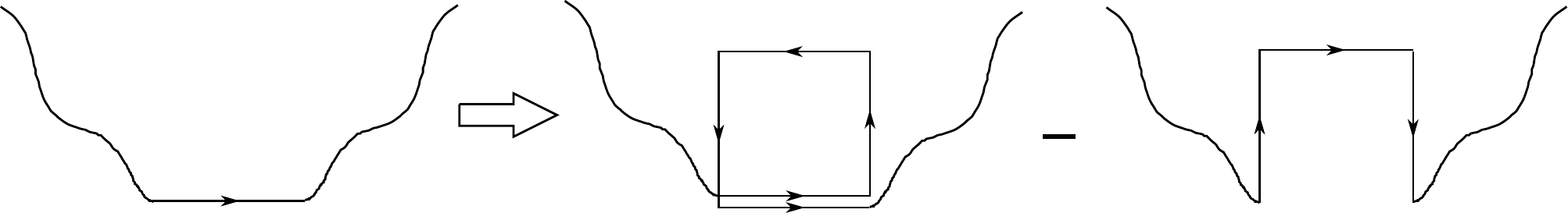}
 \caption{Intersection with the action $S*W$ term. In this case the action is simply given by a plaquette. Curvy lines represent schematically the rest of the loop.}
 \label{loopeq1}
 \end{figure}

\begin{figure}
  \centering
  \includegraphics[width=10cm]{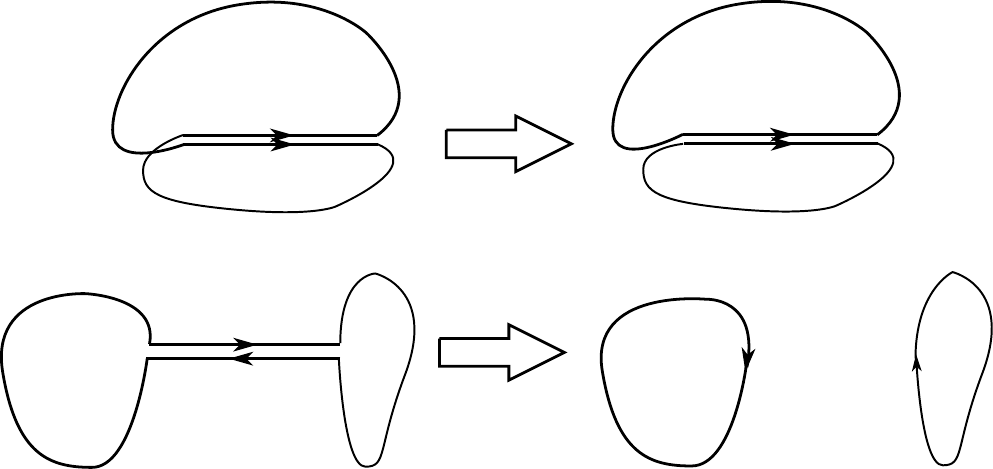}
  \caption{Self intersection terms. It only appears if the Wilson loop traverses the same link more than once (in the same or opposite directions). The first type gives a positive contribution, the second one a negative one.}
  \label{loopeq2}
  \end{figure}

 The loop equation is a direct consequence of the Schwinger-Dyson equation associated with a link of the lattice \cite{loopeqn,review1984,MakeenkoBook}. We reproduce its derivation here, first to introduce the notation, and second because for other theories the derivation will be done just by analogy to this one. 
 Consider then a point $\vec{x}$ in the lattice and a given link $\mu=0,1,2,3$, and perform the following change of variables 
\beq
U_\mu \rightarrow (1+i\epsilon)U_\mu, \ \ U^\dagger_\mu\rightarrow U^\dagger_\mu (1-i\epsilon), \ \ \ \epsilon^\dagger=\epsilon, \ \ \tr \epsilon=0\ .
\label{a11}
\eeq
The variation of the action is
\beq
 \delta S = -\frac{N}{2\lambda} \sum_{\pm\nu\neq\mu} 
 \left( i\epsilon_{ab} W^{ab}\{\mu\nu\bmu\bnu\} - i \epsilon_{ab}W^{ba}\{\nu\mu\bnu\bmu\}\right) \ ,
 \label{a12}
\eeq
where $W^{ab}\{\mu\nu\bmu\bnu\}$ indicates a Wilson line made of four links starting and ending at $\vec{x}$ and following the directions $\{\mu\nu\bmu\bnu\}$ where the bar indicates that the link is traversed in the opposite direction as $\mu$. Let us perform this change of variables in the integral
\beq
\int \cD U\ \delta_\epsilon\left(e^{-S}\ W^{ab}_{\vec{x}}\{\mu,\hat{\cC}\}\right) = 0\ ,
\label{a13}
\eeq
where $W^{ab}_{\vec{x}}\{\mu,\hat{\cC}\}$ indicates a Wilson line starting at point $\vec{x}$ in direction $\mu$ and then coming back to $\vec{x}$  along some given path $\hat{\cC}$ that in principle may pass again through the same link $\vec{x},\mu$ in the positive or negative direction. Since the change of variables does not change the value of the integral, the variation vanishes which can be expressed in the usual form 
\beq
\langle -\delta_\epsilon S\  W^{ab}_{\vec{x}}\rangle + \langle \delta_\epsilon W^{ab}_{\vec{x}} \rangle =0\ .
\label{a14}
\eeq
Explicitly
\beqa
&& \frac{iN}{2\lambda} \langle \sum_{\pm\nu\neq\mu}\left( \epsilon_{cd} W^{dc}\{\mu\nu\bmu\bnu\}-\epsilon_{cd}W^{dc}\{\nu\mu\bnu\bmu\}\right)
 W^{ab}_{\vec{x}}\{\mu,\hat{\cC}\}\rangle 
  + i \epsilon^{ad}\langle W^{db}_{\vec{x}}\rangle \nonumber \\
&& + \langle \sum_{j_+=1}^{n_+} i W^{ac}_{\vec{x}\vec{x}^+} \epsilon^{cd} W^{db}_{\vec{x}^+\vec{x}} \rangle 
-\langle \sum_{j_-=1}^{n_-} i W^{ac}_{\vec{x}\vec{x}^-} \epsilon^{cd}W^{db}_{\vec{x}^-\vec{x}} \rangle =0\ ,
\label{a15}
 \eeqa
 where the terms in the last line come from the possibility that the path goes through the same link again either in the same direction ($n_+$ times), or opposite direction ($n_-$ times). These terms are call self-intersection terms but notice that self-intersection in this context means that the loop goes through the same link more than once (in either direction) and not merely through the same vertex.
 Since this identity is valid for any traceless hermitian $\epsilon$ we conclude ($\epsilon_{cd}A^{dc} = 0 \Rightarrow A^{cd}=\delta^{cd}\frac{1}{N} A^{aa}$)
 \beqa
&& \frac{N}{2\lambda}\langle\sum_{\pm\nu\neq\mu}W_{\vec{x}}^{dc}\{\mu\nu\bmu\bnu\}W^{ab}_{\vec{x}}\{\mu,\hat{\cC}\}-W_{\vec{x}}^{dc}\{\nu\mu\bnu\bmu\}W^{ab}_{\vec{x}}\{\mu,\hat{\cC}\}\rangle + \delta^{ac}\langle W^{db}_{\vec{x}}\rangle \non\\
&& +  \langle \sum_{j_+=1}^{n_+}W^{ac}_{\vec{x}\vec{x}^+} W^{db}_{\vec{x}^+\vec{x}}\rangle 
 -\langle \sum_{j_-=1}^{n_-}W^{ac}_{\vec{x}\vec{x}^-} W^{db}_{\vec{x}^-\vec{x}}\rangle \\
&& = \delta^{cd} \left[ \frac{1}{2\lambda}\langle\sum_{\pm\nu\neq\mu} W_{\vec{x}}\{\mu\nu\bmu\bnu\}W^{ab}_{\vec{x}}\{\mu,\hat{\cC}\}- W_{\vec{x}}\{\nu\mu\bnu\bmu\}W^{ab}_{\vec{x}}\{\mu,\hat{\cC}\}\rangle + 
\frac{1}{N}\langle W^{ab}_{\vec{x}}\rangle \right.\non\\
&& \left.+ \frac{1}{N} (n_+-n_-) \langle W^{ab}_{\vec{x}} \rangle \right] \ .
\label{a16}
\eeqa
Contracting both sides with $\delta_{ac}\delta_{bd}$ we get 
 \beqa
 && \frac{N}{2\lambda}\sum_{\pm\nu\neq\mu}\langle W_{\vec{x}}\{\mu\nu\bmu\bnu\mu\hat{\cC}\}
  -W_{\vec{x}}\{\nu\mu\bnu\hat{\cC}\}\rangle + N \langle W_{\vec{x}}\rangle \non\\
&&  + \sum_{j_+=1}^{n_+}\langle W_{\vec{x}\vec{x}^+} W_{\vec{x}_+\vec{x}}\rangle  -\sum_{j_-=1}^{n_-}\langle W_{\vec{x}\vec{x}^-} W_{\vec{x}_-\vec{x}}\rangle
  = \\ 
&&   \frac{1}{2\lambda}\langle\sum_{\pm\nu\neq\mu} W_{\vec{x}}\{\mu\nu\bmu\bnu\}W_{\vec{x}}- W_{\vec{x}}\{\nu\mu\bnu\bmu\}W_{\vec{x}}\rangle  
   + \frac{1}{N} (n_+-n_-+1) \langle W_{\vec{x}} \rangle \non\ ,
   \label{a17}
  \eeqa
a useful form of the loop equation associated with each link of the loop. In the large N limit, we divide both sides by $N^2$ and get
\beq
 \frac{1}{2\lambda}\sum_{\pm\nu\neq\mu} \cW_{\vec{x}}\{\mu\nu\bmu\bnu\mu\hat{\cC}\}
  -\cW_{\vec{x}}\{\nu\mu\bnu\hat{\cC}\} + \cW_{\vec{x}}
  + \sum_{j_+=1}^{n_+} \cW_{\vec{x}\vec{x}^+} \cW_{\vec{x}_+\vec{x}} 
  -\sum_{j_-=1}^{n_-} \cW_{\vec{x}\vec{x}^-} \cW_{\vec{x}_-\vec{x}} =0\ ,
\label{a18}
\eeq
 where we defined 
\beq
 \cW\{\cC\} = \frac{1}{N} \langle W\{\cC\} \rangle \ ,
 \label{a19}
\eeq
and used the large N factorization property
\beq
\frac{1}{N^2} \langle W\{\cC_1\} W\{\cC_2\} \rangle = \cW\{\cC_1\}\, \cW\{\cC_2\} + \cO(\frac{1}{N^2})\ .
\label{a20}
\eeq
 The different terms in the result have a very simple graphical interpretation as seen in figs.(\ref{loopeq1},\ref{loopeq2}).
The link $\vec{x},\mu$ appears in the action in several plaquettes. Each of these plaquettes is connected to the loop as in the figure, when the orientations are opposite, we include a minus sign. Also the action comes with a coefficient $\frac{N}{2\lambda}$. 
 Summing over all links we get the loop equation that we schematically write as
\beq
 -\frac{1}{N L} S*\cW + \cW + \frac{1}{L}\sum_{i} 
 \s_i \cW_{1i}\cW_{2i} =0 \ ,
 \label{a21}
\eeq
where $L$ denotes the length of the loop, $S*\cW$ denotes all possible intersections between the loops appearing in the Wilson loop (at a fixed position) and those appearing in the action. The last term is a sum over all self-intersections with a sign $\s_i$ depending on the orientation of the intersection and $\cC_1$ and $\cC_2$ denote the two loops in which the original loop splits when reconnecting at that self-intersection (see fig.\ref{loopeq2}). In this form the loop equation is valid for any action given by a sum of Wilson loops. Therefore, we can give a linear combination of loops as the action and the reconnection procedure determines completely the loop equation without any reference to matrices, gauge invariance etc. 

\begin{figure}
  \centering
  \includegraphics[width=10cm]{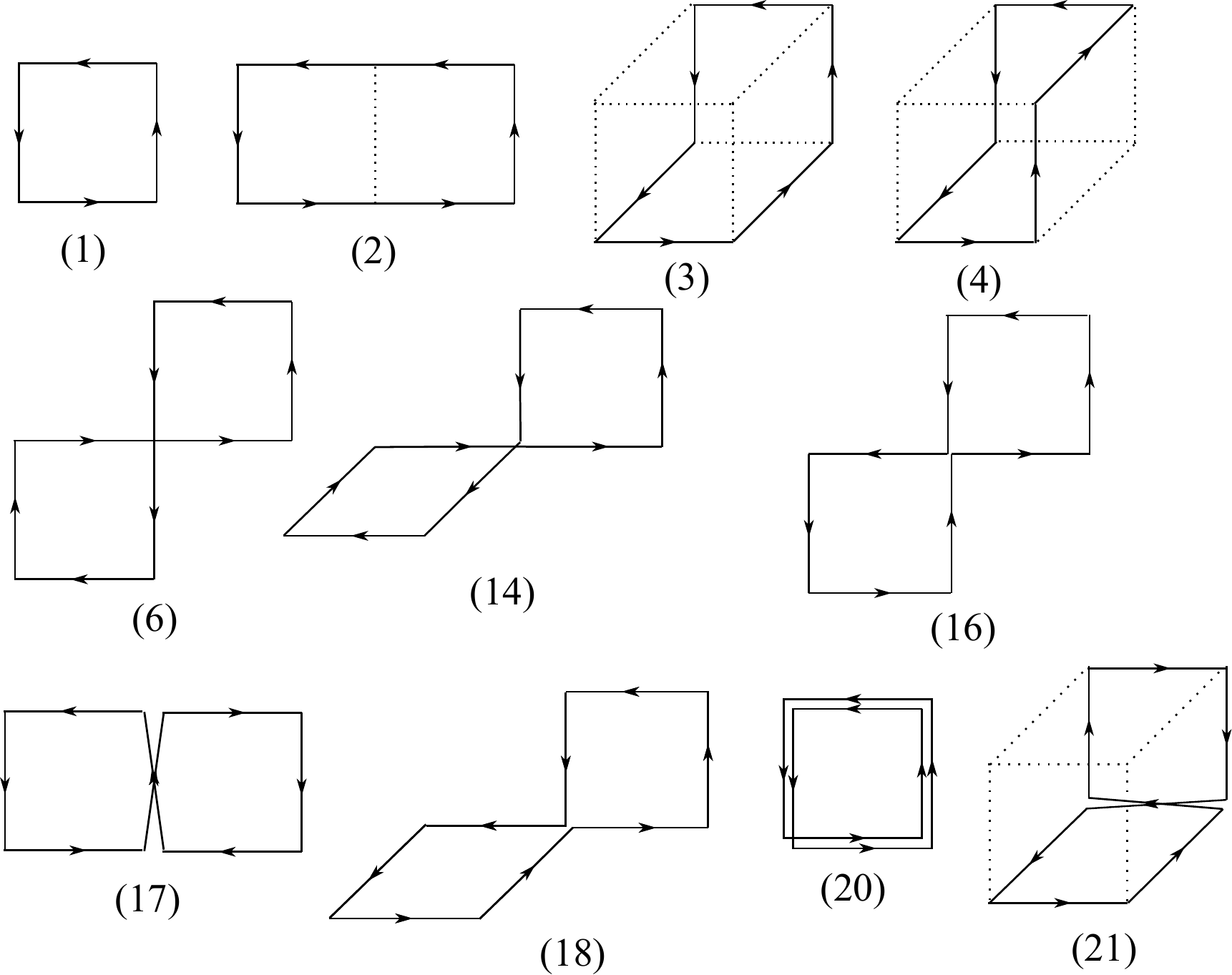}
  \caption{We construct a numbered list of Wilson loops up to translations, rotations and cyclic permutations. Some examples are in the figure.}
  \label{fig:WL}
  \end{figure}

For mathematical manipulations it is convenient to enumerate the loops in a list, where we eliminate redundancy due to rotations, translations, cyclic permutations and opposite orientations. The first few elements of the list are in fig.(\ref{fig:WL}). 
Then we can write the loop equation in the form
\beq
 \mathbb{K}_{i\rightarrow j} \cW_j + 2\lambda \cW_i + 2\lambda \mathbb{C}_{i\rightarrow jk} \cW_j\cW_k = \delta_{i1}\ ,
 \label{a22}
\eeq 
where $\cW_i$ denotes the expectation value of loop $i$, $\mathbb{K}_{i\rightarrow j}$ is a matrix indicating that loop $i$ converts into loop $j$ by the reconnection procedure of the action with a weight depending on how many different ways we can get $j$ and divided by the length of the loop $L$. The tensor $C_{i\rightarrow jk}$ is the self-intersection term and indicates that Wilson loop $i$ splits into $jk$ with an appropriate coefficient. For example, for the plaquette we get
\beqa
 -\cW_0 -\cW_2 -4 \cW_3 +\cW_{17} +\cW_{20} +4\cW_{21}  + 2 \lambda \cW_1 &=& 0 \ .
 \label{a23}
\eeqa  

\subsection{Extra equations} \label{lgt8}
 
\begin{figure}
     \centering
     \includegraphics[width=10cm]{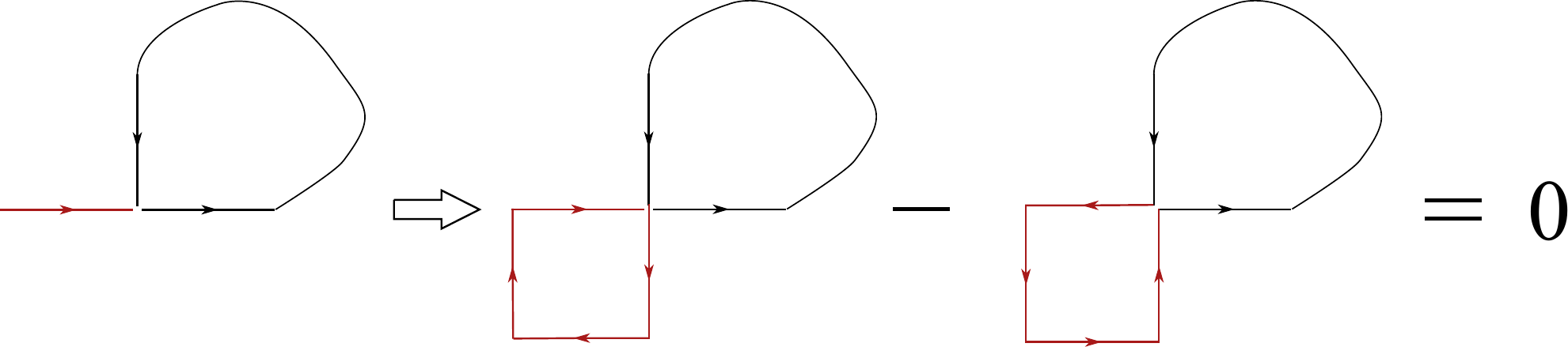}
     \caption{Loop Equation associated with a link (in red) not in the loop but sharing a vertex with the loop. The equation is linear and independent of the coupling. It can be thought as a constraint.}
     \label{loopeq3}
     \end{figure}

  From the derivation of the previous subsection it is clear that we can get more equations than just the loop equation. First we can obtain individual Schwinger--Dyson equations for each link. Given two links that do not belong to a self-intersection, the difference between their respective equations is linear and independent of $\lambda$. Other linear lambda-independent equations can be obtained from the equations associated to links that touch the loop but do not belong to it (see fig.\ref{loopeq3}). All these equations are linear and $\lambda$-independent and we denote them as constraints since do not have information on the coupling:
\beq
 B_{ij}\cW_j =0  \ .
 \label{a24}
\eeq
An example is:
\beq
 \cW_2 +\cW_6 +4\cW_{14} -\cW_{16} -\cW_{17} -4 \cW_{18} = 0\ .
 \label{a25}
\eeq
They should also be imposed since they restrict the possible values of the Wilson loops.

\section{Two dimensional lattice} \label{2D}

The two dimensional system is a well known system that can be solved exactly even in the large-N limit \cite{GW}. In this paper we use it to test numerical methods that can then be extended to the more challenging case of four dimensions.  In axial gauge, $U_0=\mathbb{I}$, the two dimensional case reduces to the single plaquette:
 \beq
  Z = \int dU e^{\frac{N}{2\lambda}\tr(U+ U^\dagger)}\ .
  \label{a26}
 \eeq
The Wilson loops can be labeled by an integer
\beq
\cW_n = \frac{1}{N}\langle \tr U^n \rangle \ .
\label{a27}
\eeq
 The large N limit was studied by Gross and Witten as well as Wadia \cite{GW} using the saddle point of an effective action for $\rho(\theta)$, the density of eigenvalues $e^{i\theta}$ of $U$ in the interval $\theta\in[-\pi,\pi]$. Later Friedan \cite{Friedan} obtained the same result using the loop equation. The result for the plaquette (and therefore the energy) is
\beq
 \cW_1=u=\left\{
 \begin{array}{lcl}
  1-\frac{\lambda}{2} & \ \ \ &\lambda\le 1 \\
  \frac{1}{2\lambda}  &       & \lambda\ge 1
 \end{array} \right.
\label{a28}
\eeq
 At $\lambda=1$ there is a jump in the second derivative of $\cW_1$ and therefore the transition is third order. Since the solution is exact we can use this system to test different methods to solve the loop equations as we do in the rest of this section. In terms of the eigenvalue density $\rho(\theta)$ (normalized to $\int_{-\pi}^{\pi}\rho(\theta)d\theta=2\pi$)
\beq
 \cW_n = \frac{1}{2\pi} \int_{-\pi}^{\pi} e^{in\theta} \rho(\theta)\, d\theta \ \ \  \Rightarrow  \ \ \ 
 \rho(\theta) = 1 +2 \sum_n \cW_n \cos(n\theta) \ ,
 \label{a29}
\eeq
where we used that the $\cW_n$ are real. Thus, the eigenvalue density is a generating function for the Wilson loops. 

\subsection{Positivity constraints, density matrix and Wilson loop entropy} \label{2D1}

 The fact that the matrices $U$ are unitary imply certain constraints that are fundamental to understand the physics and to implement the numerical methods that we describe below. We start with the observation that for any matrix $A$ with components $a_{ij}$:
 \beq
 \tr( A A^\dagger) =\sum_{ij}|a_{ij}|^2 \ge 0, \ \ \ \mbox{and} \ \ \ \tr( A A^\dagger) = 0 \ \iff \ A=0\ .
 \label{a30}
 \eeq
 Take 
 \beq
 A = \sum_{n=0}^{L} c_n U^n, \ \ \Rightarrow \ \ \ \sum_{nm=0}^{L}\bar{c}_n c_{m} \cW_{|n-m|} \ge 0, \ \ \forall c_n\ ,
 \label{a31}
 \eeq
 where we took expectation value $\cW_{|n-m|}=\langle [\tr (U^\dagger)^n U^m] \rangle$ and used that $U$ is unitary $U^\dagger=U^{-1}$. This implies that
 \beq
 \rhoM^{(L)}=\frac{1}{L}
 \left[
 \begin{array}{lllll}
 	\cW_0   & \cW_1     & \cW_2     & \ldots & \cW_{L}    \\
 	\cW_1   & \cW_0     & \cW_1     & \ldots & \cW_{L-1}  \\
 	\cW_2   & \cW_1     & \cW_0     & \ldots & \cW_{L-2}  \\
 	\vdots  & \vdots    & \vdots    & \ddots & \vdots     \\
 	\cW_{L} & \cW_{L-1} & \cW_{L-2} & \ldots & \cW_0 
 \end{array}
 \right] \succeq 0\ ,
 \label{a32}
 \eeq
 where $\rhoM\succeq 0$ indicates that $\rhoM$ is positive semi-definite. Since $\cW_0=1$ then $\tr\rhoM=1$ and therefore $\rhoM$ has properties of a density matrix. Indeed, we can write its definition as
 \beq
 \rhoM^{(L)}_{\ell\ell'} = \frac{1}{L} \sum_{ab} U^{(\ell)}_{ab} (U^{(\ell')})^*_{ab}\ ,
 \label{a33}
 \eeq 
 where $U^{(\ell)}=U^\ell$. Namely, if we take the collection of all powers of the matrix $U$ up to power $L$, the matrix $\rhoM$ traces over the color indices, the entropy $S_{WL}=-\tr \rhoM^{(L)}\ln \rhoM^{(L)}$ measures the information loss due to such tracing. Mathematically, the matrix $\rhoM^{(L)}$ is a Toeplitz matrix defined as in the eq.(\ref{a32}) or, equivalently,
 \beq
 \rhoM^{(L)}_{ij} = \frac{1}{L} \mathbb{T}[\cW_0,\ldots,\cW_L]_{ij} =\frac{1}{L} \cW_{|i-j|}\ .
 \label{a34}
 \eeq
 A useful comment is that the Toeplitz matrix is associated with the Fourier coefficients of the eigenvalue density $\rho(\theta)$ of $U$. In such case one can use Szeg\"o theorems \cite{szego} to compute limits of functions of the eigenvalues of $\rhoM^{(L)}$ by using integrals of the eigenvalue density:
\beq
 \lim_{L\rightarrow\infty} \frac{1}{L} \sum_{j=1}^{L} F(L\,\mu^{(L)}_j) =\frac{1}{2\pi} \int_{_-\pi}^{\pi} F(\rho(\theta))\, d\theta \ ,
 \label{a35}
\eeq
where $\mu^{(L)}_{j=1\ldots L}$ are the eigenvalues of $\rho^{(L)}$ and $\rho(\theta)$ is the eigenvalue density \ref{a29}. 
 
  From these constraints one can derive some simple results that are completely independent of the action that we choose:
 \begin{itemize}
 	\item  For example taking the principal minor 
 	\beq
 	\left[\begin{array}{ll}
 		1 & \cW_n \\
 		\cW_n & 1
 	\end{array}
 	\right] \succeq 0 \ ,
 	\label{a36}
 	\eeq
 	implies that all loops satisfy $|\cW_n|\le 1$ as we already know. Taking another principal minor
 	\beq
 	\left[\begin{array}{lll}
 		1 & \cW_1 & \cW_n \\
 		\cW_1 &\ 1 & \cW_{n-1} \\
 		\cW_n & \cW_{n-1} & 1
 	\end{array}
 	\right] \succeq 0\ ,
 	\label{a37}
 	\eeq
 	we obtain
 	\beq
 	(\cW_n - \cW_{n-1})^2 \le  (1-u)(1+u-2\cW_n\cW_{n-1})\ ,
 	\label{a38}
 	\eeq
 	and therefore, if $\cW_1=u=1$ then all loops are equal $\cW_{n}=1$ independently of the action! If $u<1$ we have an interesting inequality that bounds the rate of change in the Wilson loop expectation value
 	\beq
 	(\cW_n - \cW_{n-1})^2 \le  4 (1-u)\ ,
 	\label{a39}
 	\eeq
 	since $|1+u-2\cW_n\cW_{n-1}|\le 4$ because each loop satisfies $|\cW_p|\le 1$. We can then look for actions that saturate these bounds, an interesting topic that we leave for future work.
 	
 	\item Notice, from the previous item, that the bound for $\rhoM^{(L=2)}$ is $|u|\le 1$ and when we saturate it ($u=1$) we are able to compute {\em all} Wilson loops obtaining $\cW_n=1$. This is generic. Suppose we saturate the bound for $\rhoM^{(L)}$. That happens when we develop a zero eigenvalue of $\rhoM^{(L)}$, namely there is a particular set of coefficients $\hat{c}_i$ such that
 	\beq
 	\sum_{ij=0}^{L} \hat{c}^*_i \rhoM_{ij} \hat{c}_j = \langle \sum_{ij=0}^{L} \hat{c}^*_i \tr(U^\dagger)^i U^j \hat{c}_j \rangle =0\ .
 	\label{a40}
 	\eeq  
 	But this is the mean value of non-negative quantities and therefore can vanish only if it vanishes for all configurations:
 	\beq
 	\sum_{ij=0}^{L} \hat{c}^*_i\tr(U^\dagger)^i U^j \hat{c}_j =0 \ \ \ \Rightarrow \ \ \  A = \sum_{i=0}^{L} \hat{c}_i U^i =0\ ,
 	\label{a41}
 	\eeq
 	for a specific set of coefficients $\{\hat{c}_{i=0\ldots L}\}$. Notice this is a matrix equality that means we are considering only configuration that satisfy this specific constraint. Now compute
 	\beq
 	\langle \tr U^j \sum_{i=0}^{L} \hat{c}_i U^i \rangle = \sum_{i=0}^L \hat{c}_i \cW_{|i+j|} =0\ ,
 	\label{a42}
 	\eeq  
 	valid for any $j\in \mathbb{Z}$. Take the largest $r$ such that $\hat{c}_r\neq 0$ (normally $r=L$) then
 	\beq
 	\cW_{j+r} = -\frac{1}{\hat{c}_r} \sum_{i=0}^{r-1} \hat{c}_i \cW_{i+j} \ ,
 	\label{a43}
 	\eeq 
 	which is a recursion relation that allows us to compute {\em all} Wilson loops assuming that we already know the ones up to $\cW_L$. Again it should be interesting to find models that respect these equations. In the case of the Wilson action equations such as this are not exact. Nevertheless we expect a similar equation to be valid in the limit $L\rightarrow\infty$.
 	
 	\item The space of positive definite matrices is a multi-faceted convex cone. One can envision phase transitions when the minimum of the action jumps from the interior to the boundary of the cone, or between faces in the boundary. As we see later, for the case in hand the transition is of the former type. For finite $L$, the strong coupling solutions lies in the interior and the weak coupling ones at the boundary. 
 	
 	\item The matrix $\rhoM$ is positive definite for any value of $N$ and $\lambda$. For example for $N=1$ the Wilson loops are given simply by Bessel functions
 	\beq
 	\cW_n =\frac{1}{2\pi} \int d\theta  e^{in\theta} e^{\frac{1}{\lambda}\, \cos\theta} = I_n(1/\lambda)\ ,
 	\label{a44}
 	\eeq
 	implying that the matrix  $\mathbb{T}[I_0(1/\lambda),\ldots,I_L(1/\lambda)]$ is positive semi-definite for any $L$ and $\lambda$.
 \end{itemize}
 We emphasize the previous points since they translate also to higher dimensions as we discuss later.

\subsection{Effective action, numerical solution} \label{2D2}

 In \cite{GW}, the following effective action for the eigenvalue density was constructed
\beq
 S = -\frac{1}{2\pi\lambda}\int_{-\pi}^{\pi}d\theta \rho(\theta) \cos\theta + \frac{1}{4\pi^2} \dashint_{-\pi}^{\pi}\dashint_{-\pi}^{\pi} d\theta d\theta' \rho(\theta)\rho(\theta') \ln\left|\sin\left(\frac{\theta-\theta'}{2}\right)\right| \ ,
 \label{a45}
\eeq
where $\dashint$ indicates principal part. Using \cite{GR}
\beq
 \dashint_{-\pi}^{\pi} \cos n\theta \ln|\sin\frac{\theta}{2}| = -\frac{\pi}{n}, \ \ \ n\in\mathbb{Z}_{>0}\ ,
 \label{a46}
\eeq
we get, up to an additive constant, a simple effective action for the Wilson loops
\beq
 S = -\frac{1}{\lambda} \cW_1 + \sum_{n=1}^{\infty} \frac{1}{n} \cW_n^2\ .
 \label{a47}
\eeq
 Minimizing this action with respect to the $\cW_n$ trivially gives $\cW_1=\frac{1}{2\lambda}$, $\cW_{n\ge2}=0$, namely the strong coupling solution. However, for $\lambda<1$ the corresponding matrices $\rho^{L}$ are not all positive definite. Indeed, their eigenvalues are $\mu_j^{(L)}=1 + \frac{1}{\lambda} \cos\frac{\pi k}{L+1}$, $k=1\ldots L+1$ that are not all positive for $\lambda<1$. Therefore the correct problem to solve is
 \beqa
  &\mbox{Minimize}&\ \  S = -\frac{1}{\lambda} \cW_1 + \sum_{n=1}^{\infty} \frac{1}{n} \cW_n^2 \ ,\\
  &\mbox{such that}&  \rhoM^{(L)} = \frac{1}{L} \mathbb{T}[\cW_0,\ldots,\cW_L] \succeq 0\ ,
  \label{a48}
 \eeqa
for some fixed $L$. This problem has the form known as Quadratic Programming and can be converted into a problem of Semi-Definite Programming (SDP) (see \cite{sdp} and the appendix) and solved by standard packages. Using the matlab cvx package \cite{cvx} it is very easy to show numerically that for $L=10$ the results for $\cW_1=u(\lambda)$ agree perfectly with the exact answer, and even for as low as $L=6$ they are reasonably correct. Increasing further $L$ one can get more precise values for $\cW_1$ and also compute the other loops $\cW_{n=1\ldots L}$, in good agreement with the exact answer. This approach provides an excellent numerical method purely in terms of the Wilson loop expectation values and valid for all values of the coupling. Unfortuntely, in four dimensions there is no such simple action. For example, in \cite{effAction} Jevicki and Sakita derived an effective action for Wilson loops and showed that it leads to the loop equations. However, it depends on a Jacobian that is only implicitly defined. 
For that reason we turn now to the loop equation and leave further exploration of this interesting approach for the future. 

\subsection{The Loop equation and exact solution} \label{2D3}
 
 Consider the loop equation for a given loop $\cW_{n>0}$ of length $L=4n$. Every link gives rise to the same result, when connecting with a plaquette we get $\cW_{n+1}-\cW_{n-1}$. Each link self-intersects with $n-1$ other links giving $\cW_p\cW_{n-p}$ for $p=1\ldots n-1$. After dividing by the length, the loop equations becomes simply\cite{Friedan, MakeenkoBook}:
\beq
 \cW_{n+1} -\cW_{n-1} + 2\lambda \cW_n + 2\lambda \sum_{p=1}^{n-1} \cW_p \cW_{n-p} =0, \ \ \ n>0 \ ,
 \label{a49}
\eeq
 with the condition $\cW_0=1$. It is clear that, if we give a value to $\cW_1=u$, then all the other Wilson loops are fixed recursively. The equation is  very powerful but unfortunately it still leaves an infinite number of solutions, one for each value of $u$. The first observation is that $|\cW_n|\le1$ since $|\tr U^n| \le N$ because $U$ is unitary. A little numerical experimentation shows that, for $\lambda>1$ the recursion leads to divergent values of $\cW_n$ as $n$ grows, except if $u=\frac{1}{2\lambda}$. If $\lambda<1$, the recurrence diverges if $u<1-\frac{\lambda}{4}$ but is finite for $u\ge 1-\frac{\lambda}{4}$ so more constraints are needed at small coupling. Formally, following \cite{Friedan}, we can define a generating function 
\beq
 \varphi(z) = \sum_{n=1}^\infty z^n \cW_n = \frac{1}{4\lambda z}\left[z^2-1-2\lambda z+\sqrt{(z^2+1+2\lambda z)^2-4z^2(1-2\lambda u)}\right]\ ,
 \label{a50}
\eeq
 as follows directly from the loop equation. Notice that the eigenvalue density is
 \beq
  \rho(\theta) = 1 +2\sum_n \cW_n \cos(n\theta) = 1 + 2 \re[ \varphi(e^{i\theta})]\ .
  \label{a51}
 \eeq
 The condition that $|\cW_n|\le 1$ implies that $\varphi(z)$ is analytic inside the unit circle. This fixes $u=\frac{1}{2\lambda}$ at strong coupling but allows any $1-\half\lambda\le u\le 1$ at weak coupling. The extra condition that we need at weak coupling is that the eigenvalue density is non-negative
\beq
 \rho(\theta) = (1 + 2 \re[ \varphi(z)]) \ge 0\ ,
 \label{a52}
\eeq  
 which is enough to determine $\varphi$ as shown in \cite{Friedan}. The reason we repeat it here is that we wanted to emphasize the main message:
\begin{itemize}
\item There is an infinite number of solutions to the loop equation.  
\item The constraint $|\cW_n|\le 1$ greatly reduces the set of solutions, especially at strong coupling.
\item An extra condition such as eq.(\ref{a52}) is needed in order to find a unique solution. 
\end{itemize} 
 These properties can be translated into a simple numerical method which can be extended to four dimensions. Before describing it, we discuss a previous method due to Marchesini \cite{Marchesini} that produces good results at strong coupling but not at small coupling emphasizing the difficulties encountered in that region. 

\subsection{Numerical solution: Marchesini's approach} \label{2D4}

\begin{figure}
	\centering
	\subfloat[$L=10$]{\includegraphics[width=7cm]{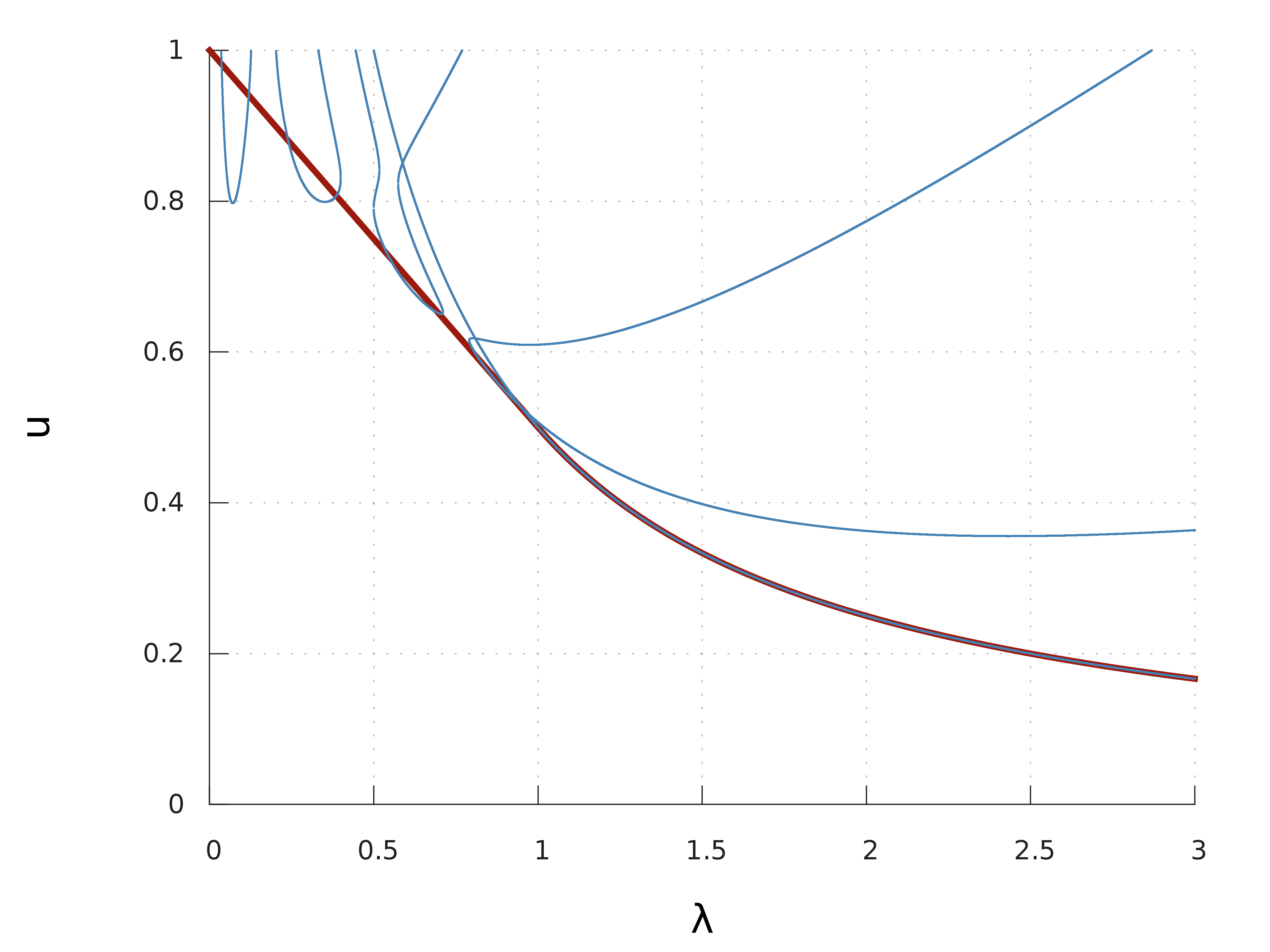}} 
	\subfloat[$L=49$]{\includegraphics[width=7cm]{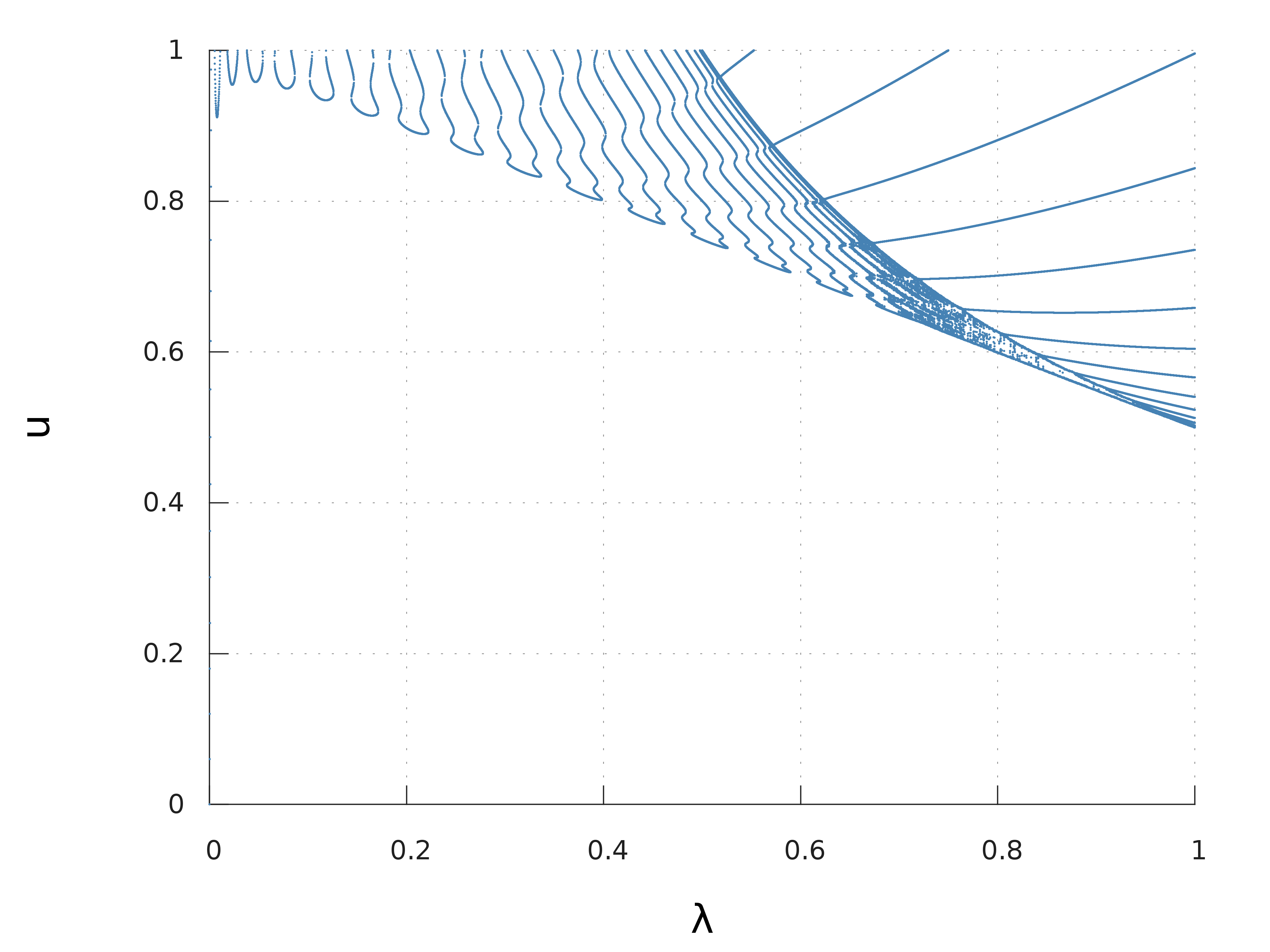}} 
	\caption{Solution to the 2D loop equation by setting to zero loops larger than $L$. In fig. (a) the strong coupling is correctly described whereas in the region $\lambda<1$ the results are not good. Going to $L=49$, fig.(b), the small coupling solution appears as an envelope of the roots but still gaps are clear. }
	\label{fig:Marchesini}
\end{figure}

 A simple numerical method to solve the loop equation was proposed and tested in two dimensions in \cite{Marchesini}. Since the number of Wilson loops is infinite, any numerical approach has to chop the set of loops. Let us assume that we consider loops up to length $4L$, namely $\cW_L$ is the last one. Given $\cW_1=u$ the loop equations for loops $\cW_{1\ldots L-1}$ determine all other loops. If we want to impose the equation for $\cW_L$ we need to know $\cW_{L+1}$. The simple proposal of \cite{Marchesini} is to set $\cW_{L+1}=0$ thus obtaining a polynomial equation for $u$:
\beq
 \cW_{L+1} = P_L(\lambda,u) = 0\ .
 \label{a53}
\eeq 
 The roots of the polynomial give the possible values of $u$. This polynomial has always a root $u=\frac{1}{2\lambda}$ that corresponds to the strong coupling solution. For small $\lambda$ other solutions appear. For example, for $L=49$ we plot the roots in fig.\ref{fig:Marchesini}. It is clear that the small coupling solution will appear in the limit of large $L$ as an envelope of the lowest roots. However, the figure does not allow us to expect a nice convergence. In \cite{Marchesini} an iterative method is proposed that converges to the lowest root and therefore it should give the correct value of $u$. However, in the same reference it is pointed out that the method requires various cut-offs and extrapolations in the weak coupling region. This suggest that its four dimensional formulation might be hard to deal with. Now we turn to a simple method that we propose in which a different polynomial whose roots provide upper and lower bounds to $u$ that quickly converges to the actual value (already $L=8$ matches the exact solution).

\subsection{Numerical approach: Bootstrap-like approach} \label{2D5}

 The main point of the numerical method is to implement as many constraints from unitarity as possible. After fixing a maximum $L$, instead of setting $\cW_{L+1}=0$ we allow it to vary under the constraint that $\rhoM^{(L)}\succeq 0$ and thus find analytical bounds for the expectation value of $\cW_1=u$. 
 
\begin{figure}
      \centering
      \includegraphics[width=14cm]{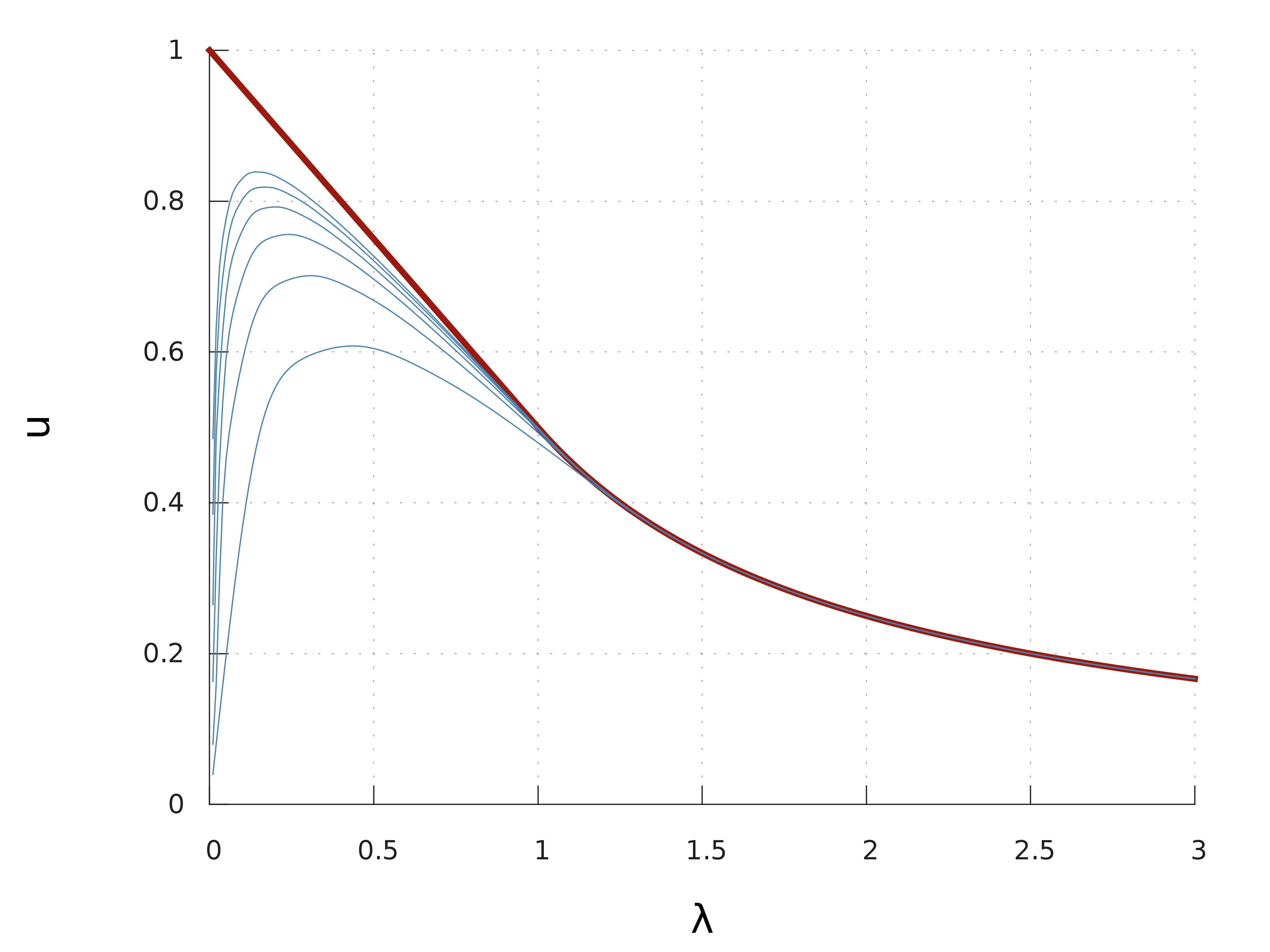}
      \caption{Solution to the 2D loop equation by imposing $|\cW_n|\le 1$ and keeping loops of length $L=11,16,21,26,31,36$.}
      \label{infimum}
\end{figure}

 Before doing that, however, we implement an even simpler idea that is to impose just the constraints $|\cW_{n=1\ldots L}| \le 1$. Starting from $L=2$ and incrementing $L$, for fixed $L$, as we vary $u$ the first loop to violate the constraint is the largest one, $\cW_L$. Therefore this is equivalent to set $\cW_L(u)=\pm 1$. From the roots of this polynomial in the region allowed by the previous step $L-1$, we choose the smallest one. This already gives better results at small coupling as can be seen in fig.\ref{infimum}. The curves are lower bounds that are continuous and converge to the exact solution as $L$ is increased (we reach $L=36$). 
 
\begin{figure}
	\centering
	\includegraphics[width=14cm]{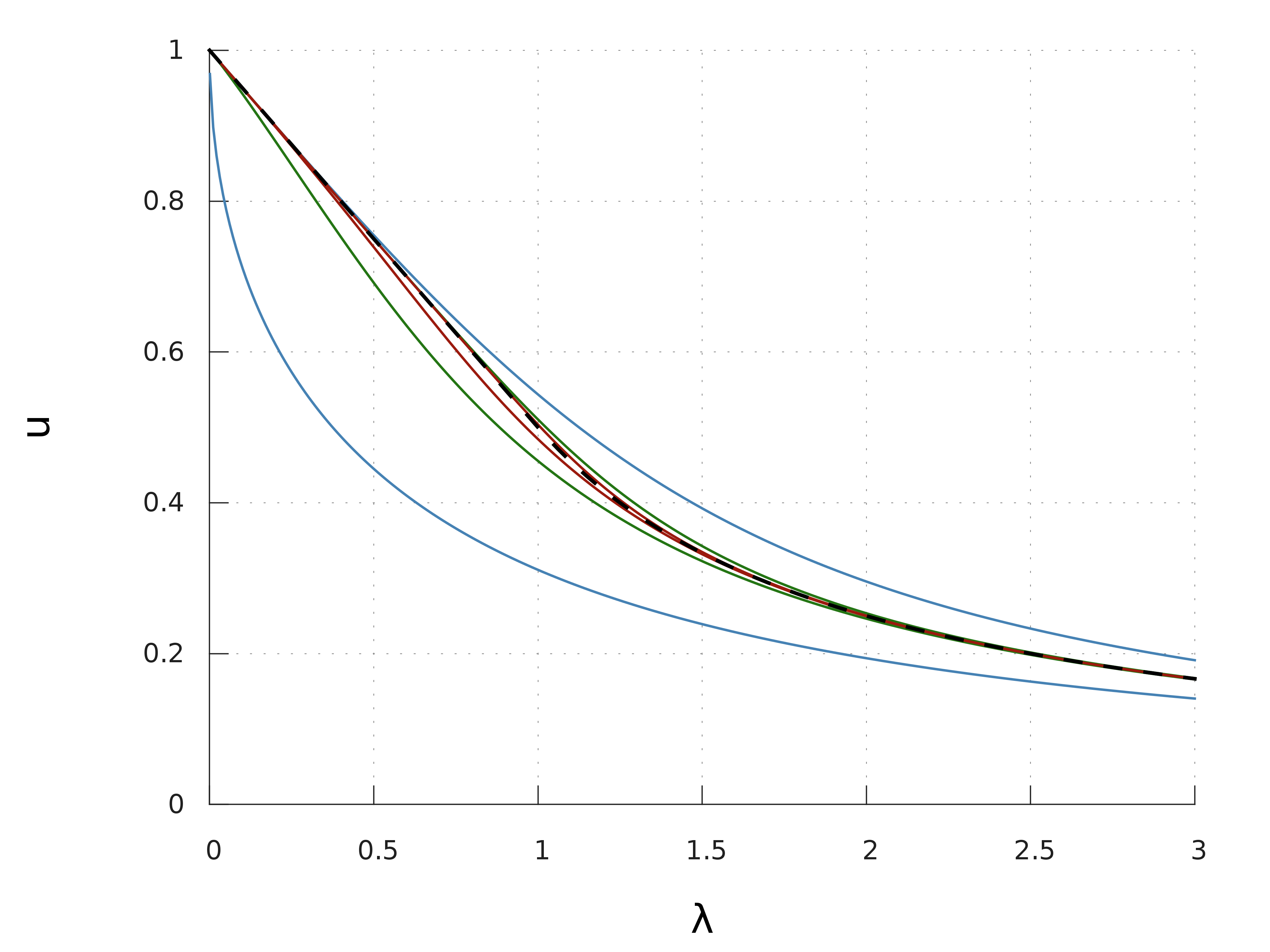}
	\caption{Upper and lower bounds to the plaquette expectation value by imposing positivity of $\rhoM^{(L)}$ for $L=4$ (blue), $L=6$ (green) and $L=8$ (red). The dashed line is the exact answer always contained between the two bounds that almost coincide for $L=8$. For small coupling the upper bound agrees with the exact answer even for $L=4$.}
	\label{fig:bootstrap2D}
\end{figure}

  Imposing all the constraints contained in $\rhoM^{(L)}\succeq 0$ is even better. In the allowed region, $\rhoM\succ 0$, all eigenvalues of $\rhoM^{(L)}$ are positive. As we vary $u$, we reach the boundary of the allowed region when an eigenvalue vanishes, namely $\det\rhoM^{(L)}=0$. This determinant is a polynomial in $u$ and therefore the roots of such polynomial determine the analytical bounds of the allowed region, for given $L$.
  Again we increase $L$ by one in each step contracting the allowed region every time. The results are displayed in fig.\ref{fig:bootstrap2D} where we can see that already for lower values of $L$ the approximation is very good. For reference we give
  \beq
   \det \rhoM^{(L=4)} = 4\lambda^2\left[u^2(u^2-1)^2-(u^2+2\lambda u-1)^2\right]
  \eeq
  The two roots of this polynomial contained in the interval $[0,1]$ correspond to the blue curves in fig.\ref{fig:bootstrap2D}.

\begin{figure}
	\centering
	\includegraphics[width=14cm]{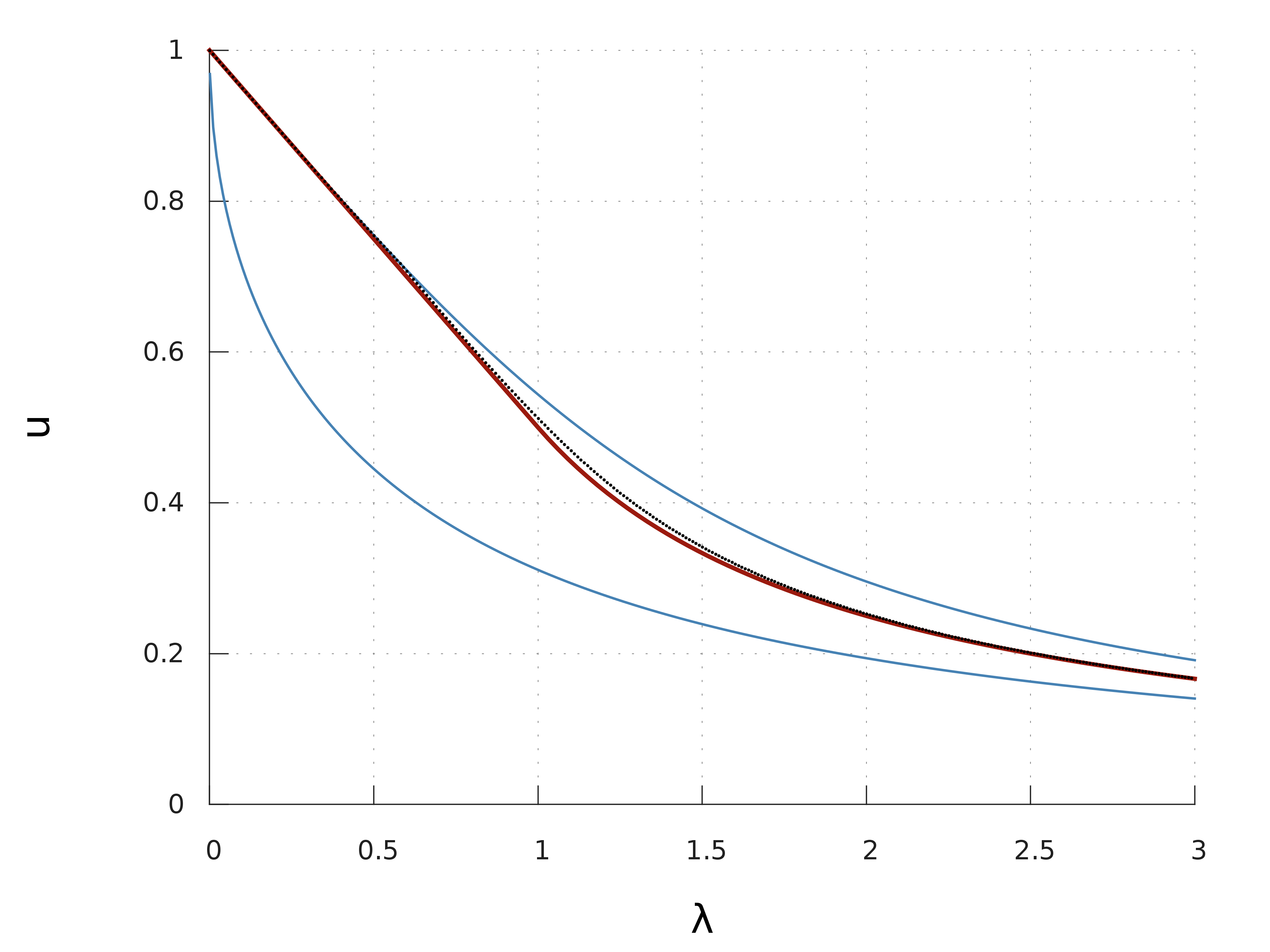}
	\caption{For $L=4$ the bounds are wide apart. However we get a good approximation (black curve) to the exact answer (red curve) using an effective action $S=-u +\half \lambda  \tr \rhoM^{(L)}\ln \rhoM^{(L)}$ }
	\label{fig:bootstrap2D2}
\end{figure}
 
\subsection{Small coupling expansion} \label{2D6}
 
 At small coupling we can use the approximation
\beq
 \cW_n = 1 - \lambda w_n +\cO(\lambda^2)\ ,
 \label{a54}
\eeq 
 and use the loop equation and positivity of $\rhoM^{(L)}$ to determine $w_n$. Again, the loop equation fixes all loops except $\cW_1=u$, namely $w_1$ in this case. Expanding the determinants of $\rho^{(L)}$ at small $\lambda$ and keeping the lowest non-vanishing term gives:
\beqa
 \det \rho^{(3)} &=& 8 \left(w_1-\half\right) \lambda^2 + \cO(\lambda^3)\, \ge 0 \ \ \Rightarrow \ \ w_1\ge \half \\
 \det \rho^{(6)} &=& - 8 \left(w_1-\half\right)^3  (2\lambda)^{10} + \cO(\lambda^{11})\, \ge 0 \ \ \Rightarrow \ \ w_1\le \half\ ,
 \label{a55}
\eeqa
 which then implies $w_1=\half$ as we know from the exact result. Higher orders should vanish as can be obtained by considering larger values of $L$. It is absolutely clear then that a systematic expansion in $\lambda$ for small coupling can only be achieved by using the $\rhoM$ matrices. The loop equation alone is not enough to fix this expansion.  

\subsection{Approximation} \label{2D7}

Although the results are excellent already for small values of $L$ we can consider a relatively low value, \eg\ $L=4$ and wonder if it is possible find an approximate value of $u$ between the maximum and the minimum. This would be an approximation as opposed to the bounds that are analytical bounds. The low coupling phase is a low temperature and therefore should minimize the energy, equivalently maximize $u$. So, at small coupling we choose $u=u_{max}$ which indeed gives a very good answer, see fig.\ref{fig:bootstrap2D}.  The large coupling or large temperature phase should have large entropy. In this case there is a simple entropy we can consider, namely the entropy associated to the matrix $\rhoM^{(L)}$. Indeed, in the limit $\lambda\rightarrow 0$, all loops are given by $\cW_n=1$, $\rho_M^{(L)}$ has one eigenvalue equal to one and all the others vanish. Namely it describes a pure state. Up to gauge transformations, the matrix $U$ is the identity and we do not lose any information if we trace its powers. On the other hand when $\lambda\rightarrow\infty$ all loops vanish except $\cW_0=1$. The matrix $\rhoM$ is proportional to the identity and the entropy is a maximum. Namely, if we take traces of powers of $U$ we lose a maximum amount of information for these configurations. 
 
 Therefore, the simple proposal is to maximize $u$ for small coupling and maximize $S_{WL}$ at large coupling. The results agrees with the exact solution better than the bounds. An even better result is obtained by defining an effective free energy
\beq
 \cA = - u + c \lambda \tr \rhoM^{(L)}\ln \rhoM^{(L)}\ ,
 \label{a56}
\eeq
where $c$ is an adjustable constant of order one. We set $c=\half$ because it seems to adjust the exact answer well (see fig.\ref{fig:bootstrap2D2}) but we do not have a way to fix this constant from first principles. Of course the correct effective action is the one we gave in section \ref{2D2} but here we wanted to find a simple effective action that could be used also in higher dimensions.

\subsection{Summary} \label{2D8}

To summarize, what we learned from the simple two dimensional case is:
There is an infinite number of solutions to the loop equation but they are restricted by imposing positivity of $\rhoM^{(L)}$.
Such condition also allows the derivation of bounds independently of the action. Once the loop equation is imposed we can derive a weak coupling expansion, strict bounds on the energy and a simple approximation when considering short loops.

\section{Four dimensional lattice} \label{4D}

 In four dimensions the numerical methods are similar as in two dimensions, the main difficulty being that the number of Wilson loops grows exponentially with the length. To handle that, we developed a computer program that listed all loops up to translations, rotations, reflections and cyclic permutations of the links up to length $L=18$ although most calculations described below were done using loops up to length $L=14$. It also computes the corresponding loop equations, strong coupling expansion, and a set of $\rhoM$ matrices that have to be positive definite as explained below. Finally it provides output that can be further manipulated by computer algebra programs or standard packages such as cvx \cite{cvx} or sdpa \cite{sdpa}. Let us now briefly describe different methods that can be used to solve the loop equation in different regimes and their usefulness. 

\subsection{Strong coupling methods} \label{4D1}

 \subsubsection{Strong coupling expansion for plaquette expectation value} \label{4D2}
 
The strong coupling expansion for the Wilson loop can be done straight-forwardly using the loop equation \cite{Marchesini}. Indeed writing the loop equation as
\beq
   \cW_i  = \frac{1}{2\lambda} \delta_{i1} -\frac{1}{2\lambda}  \mathbb{K}_{i\rightarrow j} \cW_j - \mathbb{C}_{i\rightarrow jk} \cW_j\cW_k \ ,
   \label{a57}
\eeq 
we obtain a simple solution as a series expansion
\beq
 \cW_i = \sum_{\ell=1}^\infty \frac{1}{\lambda^\ell} \cW_i^{(\ell)} \ ,
 \label{a58}
\eeq
where
\beqa
 \cW_i^{(1)} &=& \half \delta_{i1} \\
 \cW_i^{(\ell)} &=& -\half \mathbb{K}_{i\rightarrow j} \cW_j^{(\ell-1)} - \mathbb{C}_{i\rightarrow jk} \sum_{\ell'=1}^{\ell-1} \cW^{(\ell')}_j\cW^{(\ell-\ell')}_k \ .
 \label{a59}
\eeqa
 We obtain the expansion for the plaquette as
\beq
 u = \frac{1}{2\lambda}+\frac{1}{8\lambda^5}+\cO(\lambda^{-6})\ .
 \label{a60}
\eeq
Up to the computed order, the result agrees with eq.(\ref{a5}) thus providing a way to validate our computer code.
Higher order terms require going to larger loops or using other methods such as character expansion \cite{Drouffe}.

\subsubsection{Iterative strong coupling numerical solution} \label{4D3}

\begin{figure}
	\centering
	\includegraphics[width=14cm]{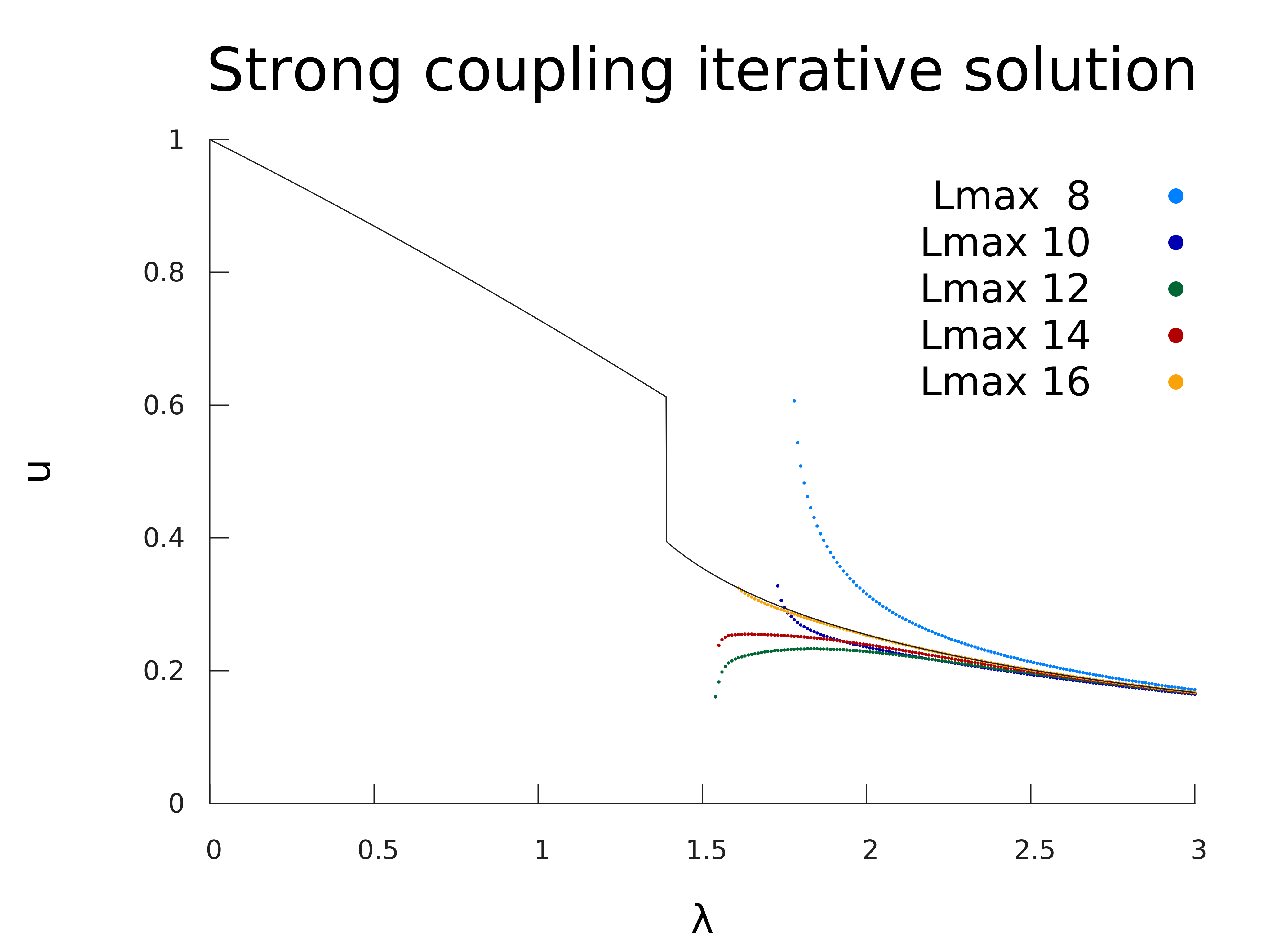}
        \caption{The results of the iterative strong coupling method are shown. Here we can see a clear improvement as loops of longer length are included in the calculation. However, all solutions are seen to diverge at $\lambda \sim 1.5$.}
	\label{fig:iterative}
\end{figure}

\begin{figure}
	\centering
	\includegraphics[width=12cm]{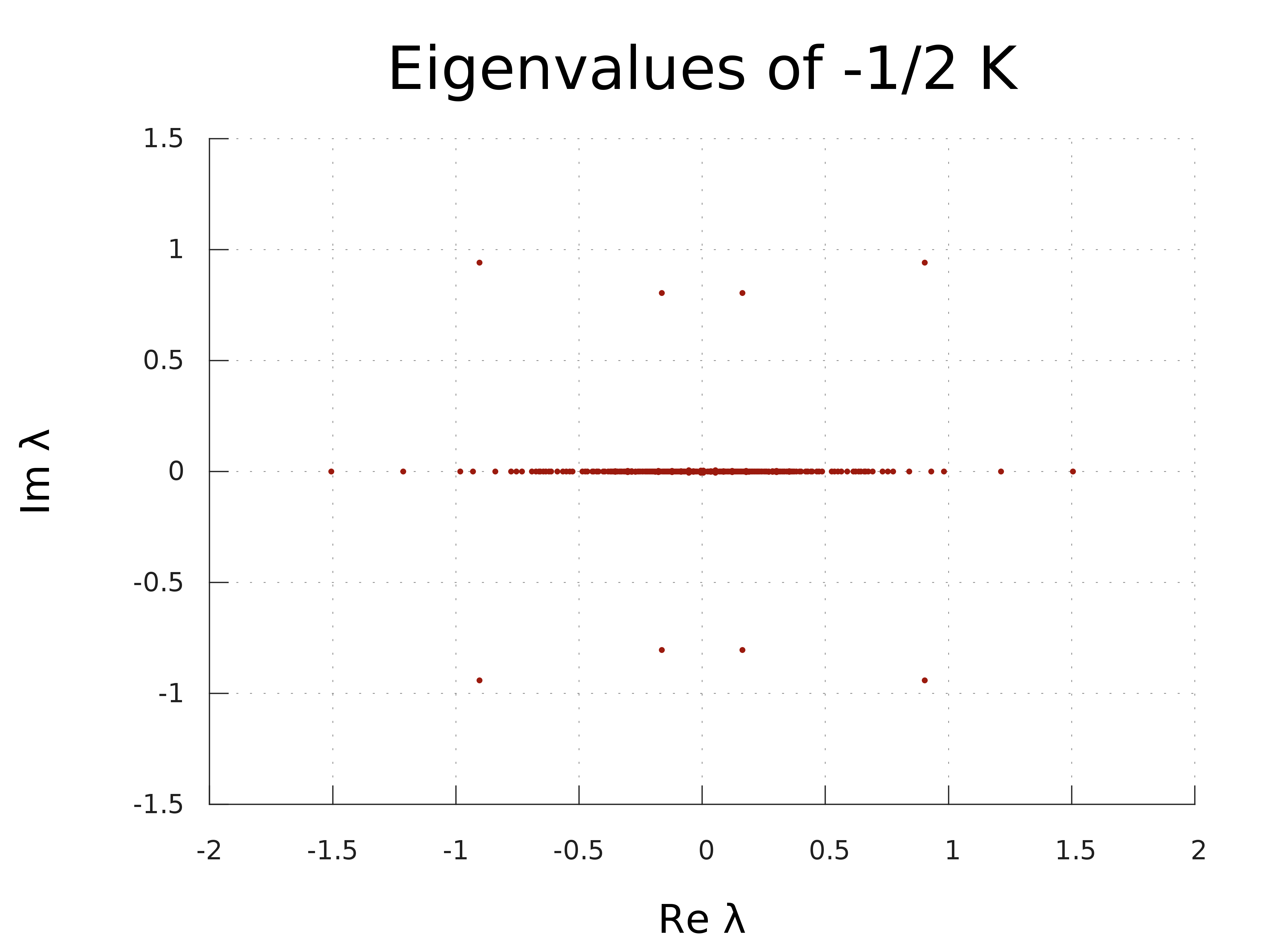}
	\caption{Setting to zero loops larger than $L=10$, we can diagonalize the operator $-\half\mathbb{K}$ (Wilson loop Laplacian). As seen in the figure, the eigenvalues seem to cluster on the real axis in an approximately interval $\lambda\in[-1.5,1.5]$  }
	\label{fig:iterativeeigenvalues}
\end{figure}

Instead of doing an analytical expansion we can do a simple numerical iteration of eq.(\ref{a57}). 
\beq
   \cW^{(n+1)}_i  = \frac{1}{2\lambda} \delta_{i1} -\frac{1}{2\lambda}  \mathbb{K}_{i\rightarrow j} \cW^{(n)}_j - \mathbb{C}_{i\rightarrow jk} \cW^{(n)}_j\cW^{(n)}_k \ .
   \label{a61}
\eeq 
As seen in fig.\ref{fig:iterative}, the results match very well the strong coupling expansion but diverge for $\lambda \lesssim 1.5$.  The reason is that, for the iterations to converge, the eigenvalues of the operator $-\frac{1}{2} \mathbb{K}$ have to have modulus less than $\lambda$. These eigenvalues are plotted in fig.\ref{fig:iterativeeigenvalues} where one can see that indeed the largest eigenvalue is $\lambda_{\mbox{max}} \simeq 1.5$.

 \subsubsection{Strong coupling solution, analytic continuation to weak coupling} \label{4D4}
 
 We can find a different iteration method to solve eq.(\ref{a22}) by doing 
\beq
   \cW^{(n+1)}_i  = \left(\mathbb{K}_{i\rightarrow j} +2\lambda \delta_{ij}\right)^{-1} \left(\delta_{j1} - \mathbb{C}_{j\rightarrow kk'} \cW^{(n)}_k\cW^{(n)}_{k'} \right)\ .
   \label{a62}
\eeq 
 Clearly the iteration is ill--defined if $\lambda$ is an eigenvalue of $-\half\mathbb{K}$. The matrix $\mathbb{K}$ is not symmetric but numerically we can still diagonalize it after truncating it by setting to zero Wilson loops larger than a certain length. For example, keeping loops up to length $L=10$, we observe that the eigenvalues of $-\half\mathbb{K}$ approximately cover an interval $\lambda\in[-1.5,1.5]$ on the real axis plus some sporadic eigenvalues in the complex plane that are likely the result of the truncation. We expect that in the limit of infinite length there is a cut on the real axis. By taking complex values of $\lambda$ one can find a simple analytic continuation of the string coupling solution to small values of $|\lambda|$ away from the real axis.  However,ß one can see that such analytic continuation is not the small coupling solution as one can actually expect on general grounds since the transition is first order. This method also shows that the divergences on the real axis are due to the fact that we truncated the loops by putting the larger ones to zero. If that were not the case we could adjust them so that the right hand side of eq.(\ref{a62}) does not contain the problematic eigenvectors of $-\half \mathbb{K}$ thus avoiding the divergences. The question arises of how should one choose the higher loops.  This takes us to the next method, namely we choose them so that that matrices $\rhoM^{(L)}$ are positive definite. 

\subsection{Positivity constraints}
 
\begin{figure}
 	\centering
 	\includegraphics[width=6.5cm]{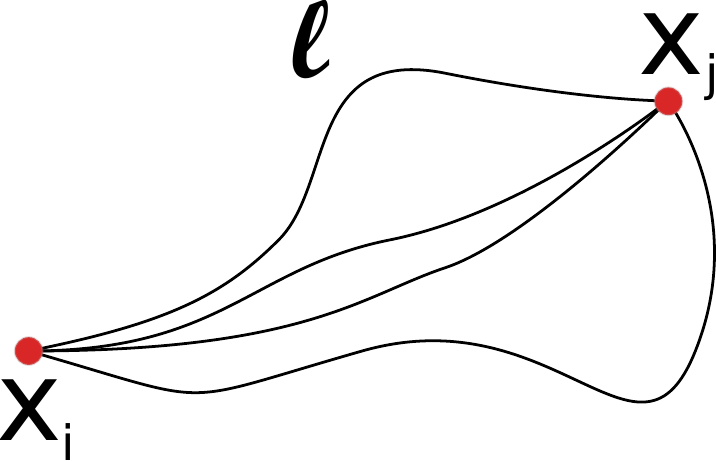}
 	\caption{A given set of open loops connecting two points in space define a matrix of closed loop expectation values $\rhoM_{\ell\ell'}=\tr\left[\left( U^{(\ell)}\right)^\dagger U^{(\ell')}\right]$ that has to be positive definite, $\rhoM\succeq 0$.}
 	\label{fig:openloops}
 \end{figure}

\begin{figure}
	\centering
	\includegraphics[width=10cm]{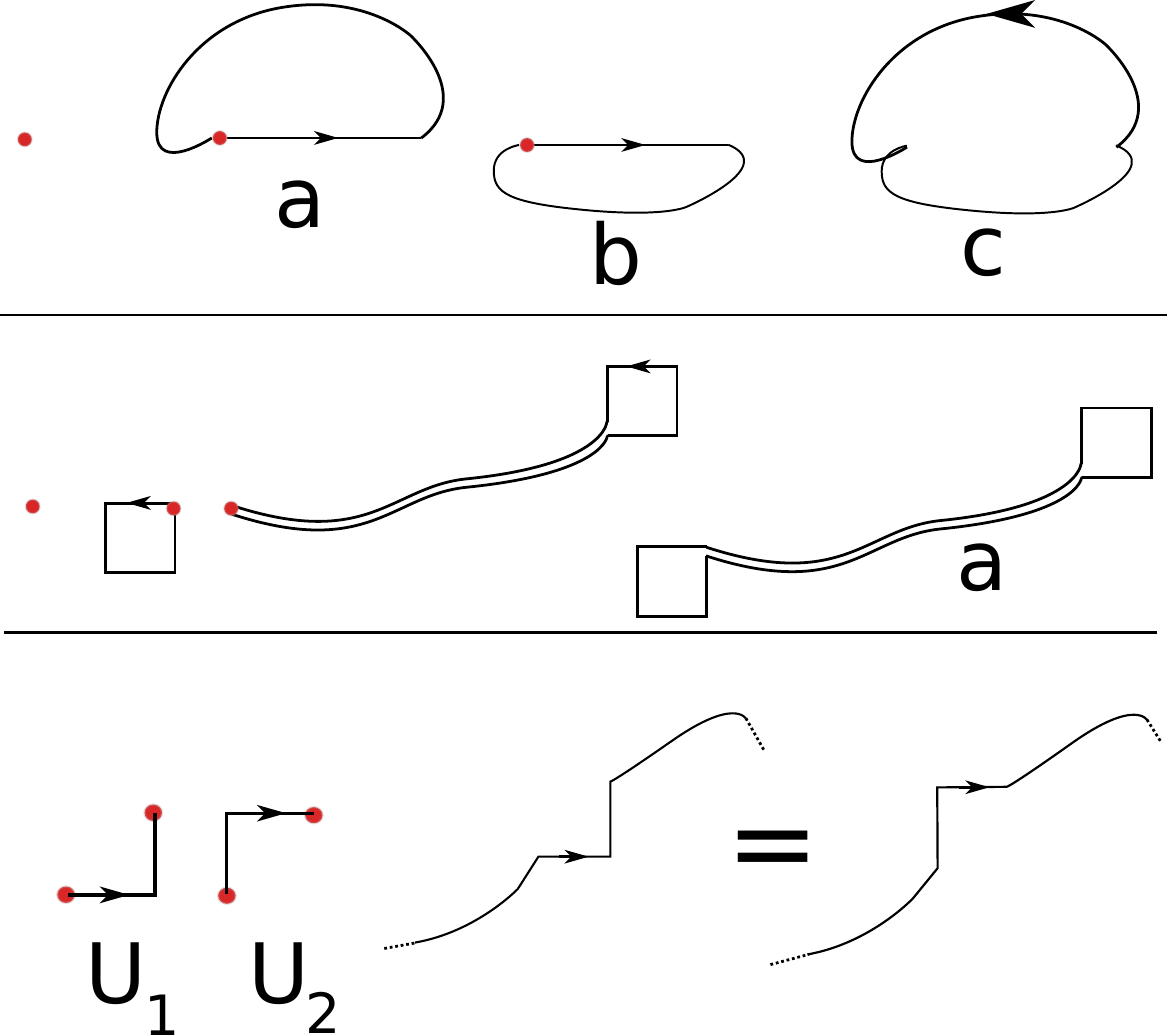}
	\caption{Three examples of applying the positivity constraints. The red dots denote a point where an open line starts and/or ends. The first one gives $ (\cW_a-\cW_b)^2 \le (1-\cW_c)(1+\cW_c-2\cW_a\cW_b)$, the second $\cW_a \ge 2 u^2-1$, where $u$ is the plaquette, and the third one implies that if $u=1$ then for all configurations the matrices $U_{1,2}$ associated with the two paths are equal $U_1=U_2$ and then all loops are equal to one.}
	\label{fig:positiveExamples}
\end{figure}

We have to construct the analogue of the matrix $\rhoM^{(L)}$ in two dimensions. The idea is very simple,
take two points $x_1$ and $x_2$ on the lattice and a set of open Wilson lines $\cC_\ell$, $\ell=1\ldots L$ going from $x_1$ to $x_2$, see fig.\ref{fig:openloops}. 
Consider an arbitrary configuration of the lattice (namely of matrices $U_\mu$ associated with each link) and compute the matrices $U^{(\ell)}$ associated with the curves $\ell$. Given an arbitrary set of coefficients $c_\ell$ we define
\beq
 A = \sum_{\ell} c_\ell U^{(\ell)}\ ,
 \label{a63}
\eeq
and since $\tr A^\dagger A\ge 0 $ for any $c_\ell$, and using that the average of a non-negative quantity is non-negative, we find that the matrix of Wilson loop expectation values 
\beq
 \rhoM_{\ell\ell'} =\langle \tr\left[\left( U^{(\ell)}\right)^\dagger U^{(\ell')}\right] \rangle\ ,
 \label{a64}
\eeq
is positive semi-definite for any set of open loops and any two points $x_{1,2}$. This is true in the continuum and in the lattice. For the lattice case we choose different pairs of points and list all possible open loops up to a certain length.  In this way we construct a set of matrices that have to be positive definite. Before going into the numerical procedure let us notice a few facts regarding this matrices that are independent of the action that we use to average the Wilson loops. 
\begin{itemize}
	\item If we take $x_1=x_2$ and two loops that start and end at $x_1$ and share the first link, see fig.\ref{fig:openloops}, we can construct the matrix
	\beq
	 \rhoM =\left(\begin{array}{ccc}
      1 &\cW_a &\cW_b \\ 
      \cW_a & 1  &\cW_c \\
      \cW_b &\cW_c &1 \\
	 \end{array}\right) \succeq 0 \ \ \Rightarrow \ \ 0\le (\cW_a-\cW_b)^2 \le (1-\cW_c)(1+\cW_c-2\cW_a\cW_b)\ ,
 \label{a65}
	\eeq  
where $\cW_c$ is the intersection of $\cW_a$ and $\cW_b$ as in the figure. In particular, if $a$ is a plaquette, $c$ is any loop that appears in the Laplacian of $b$, namely in the loop equation for $b$. 
\item Another example is a long loop $a$ made out of two plaquettes connected by a long path (fig.\ref{fig:openloops}). Cutting this loop in two as in the figure and using the same procedure we obtain 
\beq
\cW_a \ge 2 u^2-1\ ,
 \label{a66}
\eeq
where $u$ is the expectation value of the plaquette. This means that, if the expectation value of the plaquette is close to one, there are arbitrary long loops that are also close to one. 
\item Finally notice that, as in 2D, if the matrix $\rhoM$ has a zero eigenvalue, namely it is in the boundary of the allowed region, then for a given set of coefficients $\hat{c}_\ell$ in eq.(\ref{a63}) the matrix $A$ vanishes and this in turn implies an infinite set of linear equations for loops that contain the given open paths.  Namely, taking an arbitrary path $\cC_0$ from $x_j$ to $x_i$ we get  
\beq
 \langle \sum_{\ell=1}^{L} \hat{c}_\ell \tr\left[U_0 U^{(\ell)}\right] \rangle =0\ .
\eeq
As a simple example, let us show that if the plaquette is one ($u=1$) then all loops are equal to one. Indeed, take two paths that build a plaquette as in fig.\ref{fig:positiveExamples}, we get the matrix
\beq
 \rhoM =\left(\begin{array}{cc}
 	1 & u  \\ 
 	u & 1  
 \end{array}\right) \succeq 0 \ \ \Rightarrow \ \ |u|^2\le1\ ,
\label{a67}
\eeq
as expected. However, if the bound is saturated, \ie\ $u=1$ then the matrix associated with the zero eigenvalue should vanish. Namely $A=U_1-U_2=0$. Since we can use this in any loop it simply means that, for any loop, performing a move such as the one in the same figure does not change its expectation value. Since any loop can be reduced to a point by such moves, then all loops are equal to one.
\end{itemize}

\subsection{Bootstrap-like method} \label{4D5}

 The numerical method should now be clear. We list all loop equations and extra equations we described previously and that can be handled by the computer resources available. Then we construct all matrices $\rhoM$ that we can handle and look for solutions of the loop equation that satisfy the constraint $\rhoM\succeq 0$. This gives an upper and lower bound on $u$. Let us first do this analytically for the simplest example, the loop equation for the plaquette in dimension $d\ge 3$. In the notation of fig.\ref{fig:WL}, the equation is
\beq
 2\lambda u = 1+\cW_2 + 2(d-2) \cW_3 - \cW_{20} - \cW_{17} - 2(d-2) \cW_{21}\ .
 \label{a68}
\eeq 
If we want to maximize $u$, we set $\cW_2=\cW_3=1$, their maximum values and $\cW_{20}=\cW_{17}=\cW_{21}=2u^2-1$ their minimum values according to the previous subsection, eqs.(\ref{a65}) or (\ref{a66}). We then get an equation
for the maximum
\beq
 u^2 + \frac{\lambda}{2(d-1)} u -1 =0\ .
 \label{a69}
\eeq
One of the roots of this equation gives the upper bound, namely
\beq
 u\le \half\left(-\frac{\lambda}{2(d-1)}+\sqrt{4+\frac{\lambda^2}{4(d-1)^2}}\right) =u_\mathrm{max} \ ,
 \label{a70}
\eeq
a bound valid for the Wilson action in a cubic lattice of any dimension $d\ge 3$. For $\lambda\rightarrow\infty$ we get $u_\mathrm{max}\simeq\frac{2(d-1)}{\lambda}$ and for $\lambda\rightarrow 0$, $u_\mathrm{max}\simeq 1-\frac{\lambda}{4(d-1)}$. The bound has the right behavior but the coefficients are not right, as was somewhat expected from such crude bound. 
This calculation is simply an illustration that there is indeed a bound that follows from positivity of $\rhoM$ and the loop equation.
Similarly we can get a lower bound by choosing $\cW_2=\cW_3=2u^2-1$, and $\cW_{20}=\cW_{17}=\cW_{21}=1$:
\beq
 u \ge \half\left(\frac{\lambda}{2d-3}-\sqrt{4+\left(\frac{\lambda}{2d-3}\right)^2}\right) = u_\mathrm{min}\ .
 \label{a71}
\eeq
To go further we can use a numerical procedure that allows us to handle $\rhoM$ matrices of size of order $10^3\times 10^3$ and thousands of equations. The main obstacle is that the known numerical packages (see appendix) only deal with linear equations. The loop equation is non-linear but we can go around it by fixing a set of loops such that the equations become linear in the rest of the variables and then exploring the available space of such loops. In this paper we only consider the case where we fix the plaquette $u$, and therefore we have to explore only a one-dimensional space thus simplifying  
the calculation. In practice we consider loops up to a given maximum length $L$ and the actual procedure depends on $L$. Let us now consider each case. 

 \subsection{Linear case, $L_\mathrm{max}=8,10$} \label{4D6}
 
 Since we consider Wilson loops of maximum length given by $L_\mathrm{max}=8,10$ we  can only impose the loop equation associated with loops up to length $L=6$. Those loops do not have self intersections and therefore the loop equations are linear. In this case we can solve the problem using semi-definite programming in a direct way. For example using matlab and cvx \cite{cvx} (see appendix) we find the bound depicted in fig.\ref{fig:h5_800}. We already see that the results are reasonable for the maximum value of $u$ at small coupling. The minimum value of $u$ for $L_\mathrm{8,10}$ turns out to be $u_\mathrm{min}=0$ which is a correct but rather poor bound for the actual value of $u$. So, we consider now Wilson loops of larger length.
 
 \subsection{Non-linear case in one variable, $L_\mathrm{max}=12,14$} \label{4D7}

\begin{figure}
	\centering
	\includegraphics[width=14cm]{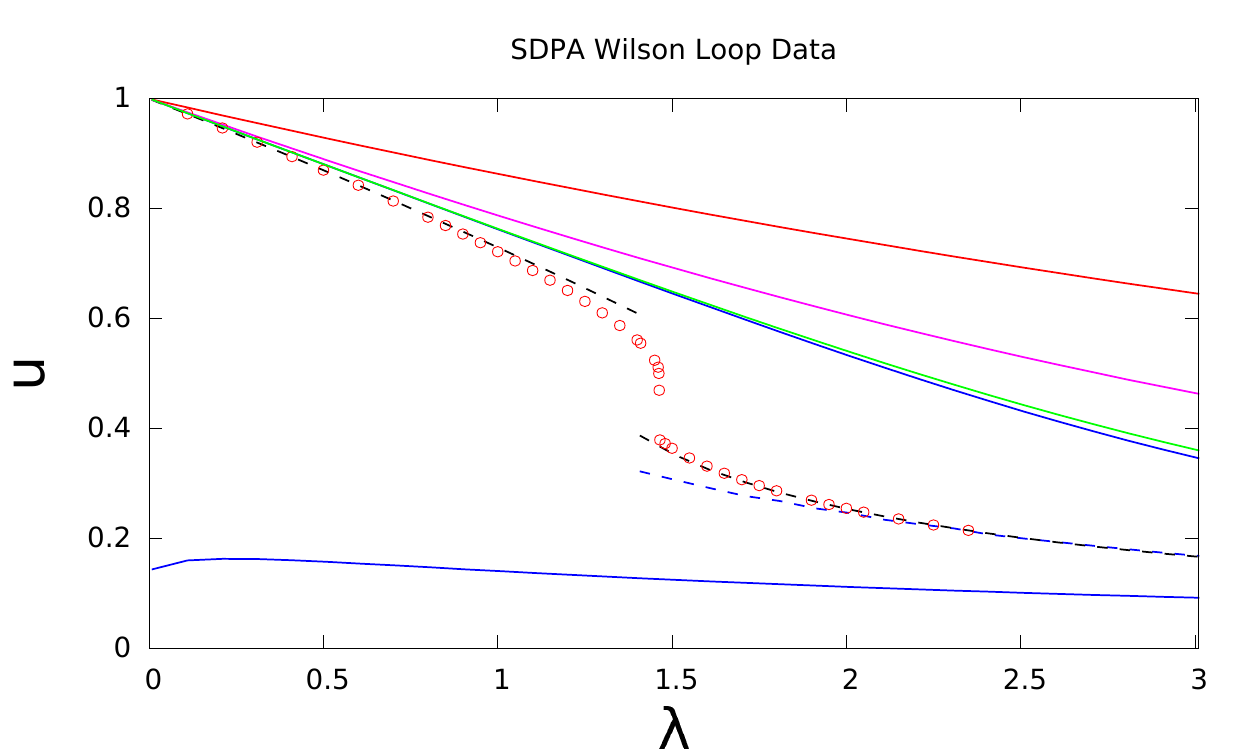}
	\caption{Expectation value of the plaquette as a function of 't Hooft coupling in four dimensions. The black lines are the small and large coupling expansion and the circles the Monte Carlo simulation. The red and purple curves are the upper bounds from $L=8,10$ (see main text). The green line is the upper bound for $L=12$ and the blue lines are the upper and lower bounds for $L=14$, the best bounds we can presently get. At small coupling the upper bound is a reasonable approximation to the plaquette. At strong coupling the blue dashed curve is a good approximation and correspond to the maximum of $t$ for $L=14$ (see main text). }
	\label{fig:h5_800}
\end{figure}

In this case we impose the loop equation up to length $L=10$. Some of those loops self-intersect and we cannot use semi-definite programming directly. However, the self-intersection splits the loops into two whose total length is less or equal than $L=10$ and therefore one of them at least has to be a plaquette. For that reason we propose a different semi--definite programming problem. We fix the value of $u$ and define a new matrix $\rhoM$ as 
\beq
 \rhoM(t) = \rhoM - t \mathbb{I}\ .
 \label{a72}
\eeq
Now we maximize the value of $t$ by allowing the loops other than the plaquette to take arbitrary values compatible with the loop equation. When the procedure finalizes, the value of $t$ is equal to the lowest eigenvalue of $\rhoM$ and it is the largest lowest eigenvalue that can be found for that fixed value of $u$.   
Thus, if $t_{max}<0$ it is simply not possible to choose the other loops such that the loop equation is satisfied and the matrix $\rhoM\succeq 0$. Therefore this value of $u$ is not allowed. In that way we can sweep the allowed values of $u$ and determine the bounds on $u$ and therefore on the energy. The results are depicted in fig.\ref{fig:h5_800} where the matrix $\rhoM$ was truncated to an $800\times 800$ size. 
The bounds are not close to each other as they were in two dimensions. At small coupling we minimize the energy and therefore choose the maximum value of $u$. Following the ideas discussed in section \ref{2D7} for the two dimensional case, for large coupling we should maximize the entropy of the matrix $\rhoM$. This is not an SDP problem and therefore we do a further approximation. Consider the value of $u$ where $t_{max}(u)$ has a maximum. One can associate such point with a large value of entropy since increasing the minimum eigenvalue, with the trace being fixed, tends to make all eigenvalues similar. We are going to take such value of $u$ as the best guess of the one that maximizes the entropy. This is depicted in fig.\ref{fig:h5_800} where we see that it is quite a good approximation at strong coupling. 

\subsection{Small coupling expansion} \label{4D8}

 As in section \ref{2D6}, we can do a small coupling expansion by linearly expanding all loops as
\beq
 \cW_n = 1 - \lambda w_n + \cO(\lambda^2)\ ,
\eeq
 and finding bounds on $w_n$. Using sdpa for loops up to length $L=12$ we find for the plaquette $w_1\ge 0.238$ and for $L=14$ $w_1\ge 0.2495$ in agreement with the value $w_1=\frac{1}{4}$. Unfortunately for this length we do not find a minimum value of $u$ and we cannot fix $w_1=\frac{1}{4}$ as we were able to do in two dimensions.  

\subsection{Lattice simulation} \label{4D9}

Our Monte Carlo simulation used the Metropolis Algorithm to produce an ensemble of uncorrelated configurations using the partition function in eqs.(\ref{a1}), (\ref{a2}):
\beq 
Z = \int\prod_{x,\mu} d U_\mu(x) e^{-S[U]} \ .
\label{a73}
\eeq 
In order to ensure that the results were properly thermalized we allowed the simulation to run for 20000 updates before saving configurations. Each update consisted of 10 ``hits'' on each link. A hit
is one attempt to move the link through the phase space. After thermalization, one configuration every hundred was saved until a total of 300 configurations were collected. Binning showed that the correlation time
of the system was around 400 updates between saves. Error was calculated via the bootstrap method and we found that for Wilson Loops up to length 10 the error in the expectation value was around $10^{-5}$ for lattices of size $8^4$. The
simulation was programmed using CUDA on GeForce GTX 980s. The expectation value of each Wilson loop was computed by averaging over configurations and over all possible positions and rotations of the loop. This allowed us
to check the loop equations validity explicitly through the Monte Carlo Simulation.
 The finite N loop equations (\ref{a17}) were checked to be valid for $3\le N\le 10$. We want to emphasize 
that for the equation to be valid for the $SU(N)$ case the right--hand side of eq.(\ref{a17}) is necessary. 
 Furthermore, the large-N loop equations (\ref{a18}) were satisfied up to order $\frac{1}{N^{2}}$ corrections. 

\begin{figure}
	\centering
	\includegraphics[width=14cm]{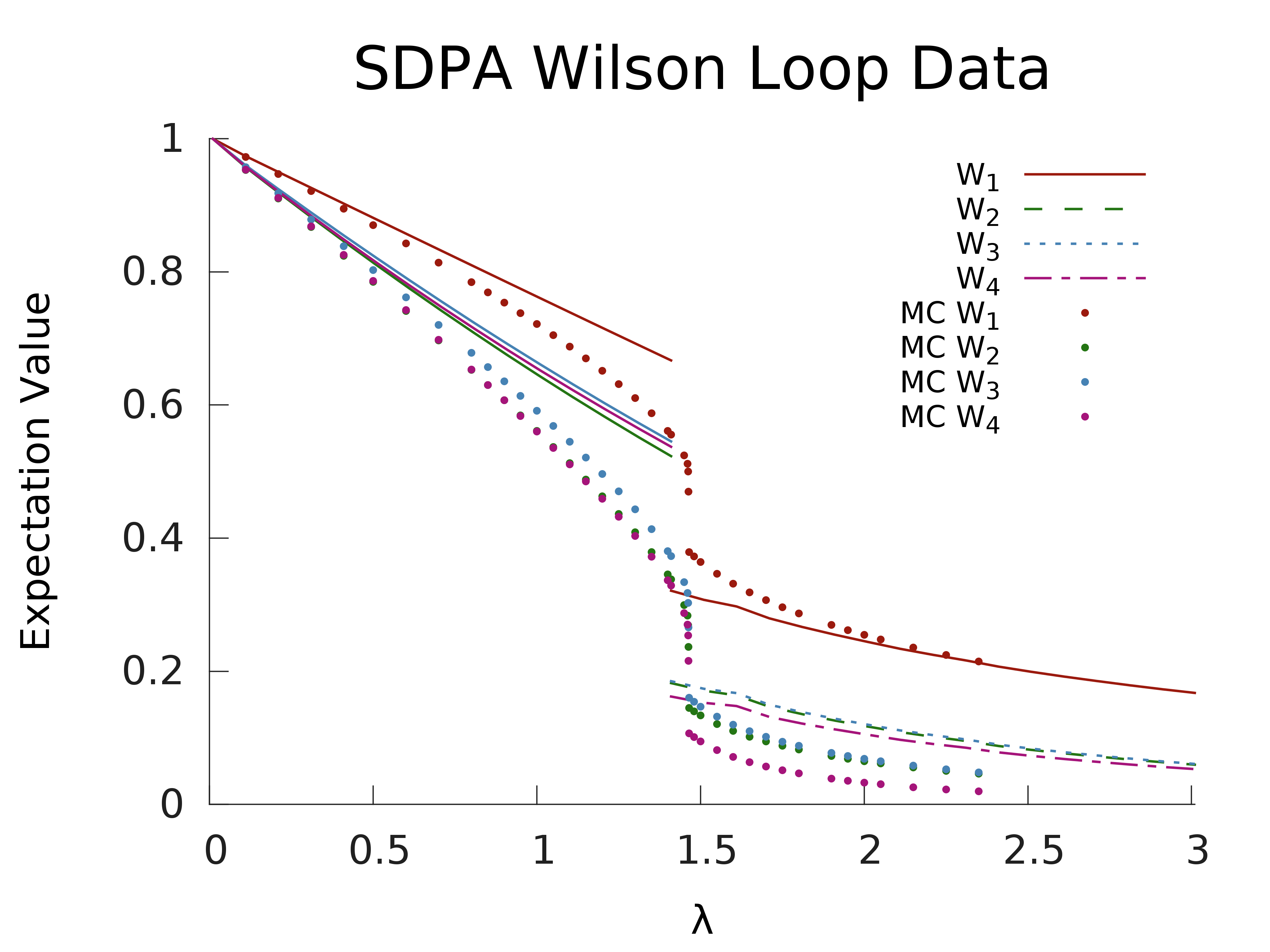}
	\caption{Expectation value of loops  $u=\cW_1$, $\cW_2$, $\cW_3$, $\cW_4$ (see fig.\ref{fig:WL}) are displayed as a function of the 't Hooft coupling $\lambda$.  }
	\label{fig:sdpr}
\end{figure}

Finally, we can compare the simulation and the bootstrap results for loops other than the plaquette. In fig.\ref{fig:sdpr} the results for loops $\cW_2$, $\cW_3$, $\cW_4$ (see fig.\ref{fig:WL}) are displayed. The agreement is reasonable at  small coupling $\lambda\lesssim 0.5$ if we used the solution that minimizes the Energy (maximizes $u$), in agreement with our previous discussion. For larger values of the coupling but below the transition the agreement is qualitative in the sense that the larger loops have smaller values than the plaquette. In the strong coupling phase, we choose the solution that maximizes $t$ but the previously found agreement (fig.\ref{fig:h5_800}) for the plaquette does not extend to the other loops. Clearly, the well-known strong coupling expansion is still the best method in this region. 

\section{$\mathcal{N}=4$ SYM} \label{N4}

The case of $\N{4}$ SYM is particularly important since it has a dual description as a string theory \cite{malda}. In the context of its relation to string theory and more precisely with the AdS/CFT correspondence, the loop equation has been formulated and studied in \cite{loopeqStringsN=4}. Studying the theory in the lattice should allow a different method of computation and
possible strong coupling calculations based on the gauge theory side of the correspondence. In the rest of this section we briefly describe how the ideas we developed for pure YM could be implemented but, for concrete calculations, at the moment we restrict ourselves to the bosonic sector
leaving the study of the full theory for future work.

\begin{figure}
\centering
\includegraphics[width=12cm]{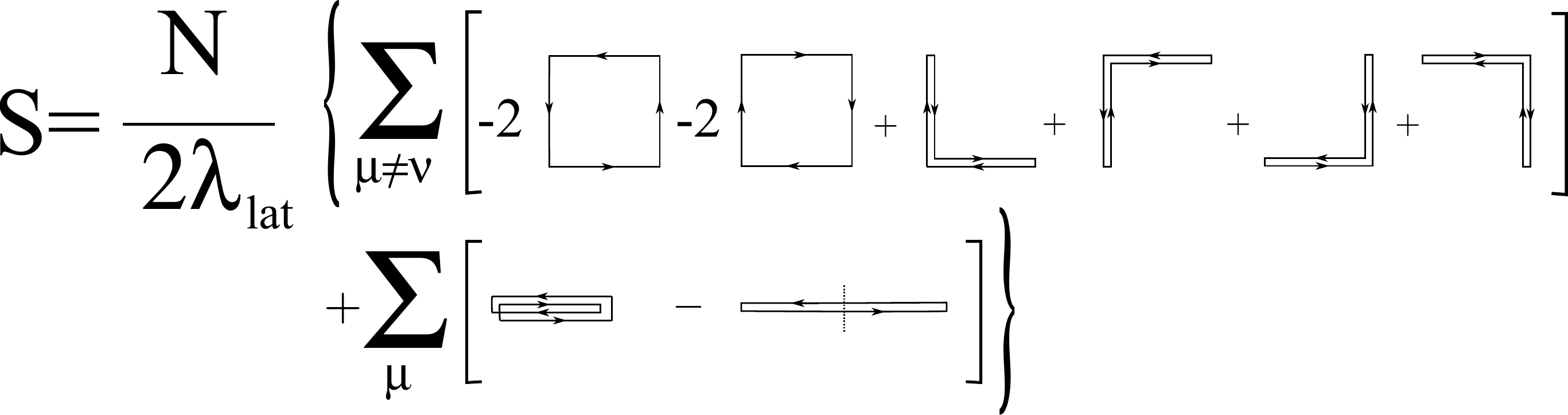}
\caption{The Lattice action for \N{4} SYM can be written in terms of Wilson loops, the figure shows the bosonic sector that we use in the main text. The indices $\mu,\nu=1\ldots 5$ since the $A4^*$ lattice has five fundamental vectors.}
\label{fig:N4action}
\end{figure}

 A lattice theory that has the correct continuum theory without the need for fine-tuning is described in \cite{Catterall:2009it}.
Here we use that formulation but follow the notation found in the paper \cite{Catterall:2013roa}. For brevity we do not 
explain details and refer the reader to that work for explanation of the notation and properties of the theory. 
The main property is that such formulation preserves one twisted \cite{Marcus} scalar supercharge and possesses BPS Wilson loops although more restricted than the continuum theory. For our purposes another important property is that the action can be formulated entirely in terms of generalized Wilson loops, namely using loops with fermionic links and/or sites. In this way, the form of the loop equation given in (\ref{a21}) is valid using the  appropriate supersymmetric action \cite{Catterall:2013roa}. 
However, we have not worked out the correct generalization of the $\rhoM$ matrices to the fermionic sector and therefore here we restrict ourselves to the bosonic sector described by the simpler action
\beq
S = \frac{N}{2\lambda_{lat}} \sum_x \Tr\Bigg( 
\F{ab}^\dagger(x)\F{ab}(x) + \frac{1}{2}\bigg(\Dbm{a}\U{a}(x)\bigg)^2
\Bigg)\ .
\label{a74}
\eeq
This action, thought as a linear combination of Wilson loops is depicted in fig.\ref{fig:N4action}. In order to obtain the loop equations we must vary individual links. However, the scalar and the gauge fields are twisted together which means that the matrices $\U{a}(x)\in GL(N,\mathbb{C})$ and therefore the links and their daggers must be treated independently:
\begin{align}
  \U{a}(x) &\Rightarrow  (1 + i \epsilon_a(x))\U{a}(x) \\
  \Ub{a}(x) &\Rightarrow  \Ub{a}(x)(1 - i \bar{\epsilon}_a(x)) \ .
  \label{a75}
\end{align}
This implies that in eq.(\ref{a21}), the intersection of the action with the loop is non-zero only for links oriented in the same direction, the same is true for a loop self-intersection. Since the gauge group is $U(N)$ the $\epsilon$'s do not have the constraint that they have to be traceless simplifying the Loop Equations. 
 The matrices $\rhoM$ can be constructed similarly since we associate the hermitian conjugate of $\U{a}$ to a link traversed in the opposite direction. Notice that in this case the property that a backtracking path can be eliminated is no longer true. In fact such backtracking paths correspond to the insertion of scalar fields.  Now we check the loop equation using a numerical simulation and leave the full exploration of the bootstrap method for future work.

\subsection{Monte Carlo Simulation} \label{N43}
We used the parallel code developed by Schaich and DeGrand \cite{Schaich:2014pda}, based on
the previous work by Catterall and Joseph \cite{Catterall:2011cea} to simulate the
latticized Euclidean theory. The code allows for a simple way to reduce to the bosonic 
sector, \ie\ to the six scalars and four gauge fields that in this formulation live on the
links. The lattice is taken to be an $A_4^*$ lattice which contains the permutation group $S_5$, 
the largest finite subgroup of the four dimensional rotation symmetry. In order to preserve
a subgroup of the SUSY Algebra, the Euclidean-Lorentz and R symmetry groups are twisted
into $SO(4)_E \otimes SO(6)_R \rightarrow SO(4)' \otimes U(1)$. With this twisting 
the lattice theory is invariant under one supercharge out of the full sixteen. The continuum 
limit should restore the full sixteen supercharges without fine-tuning.

\begin{figure}
	\centering
	\includegraphics[width=10cm]{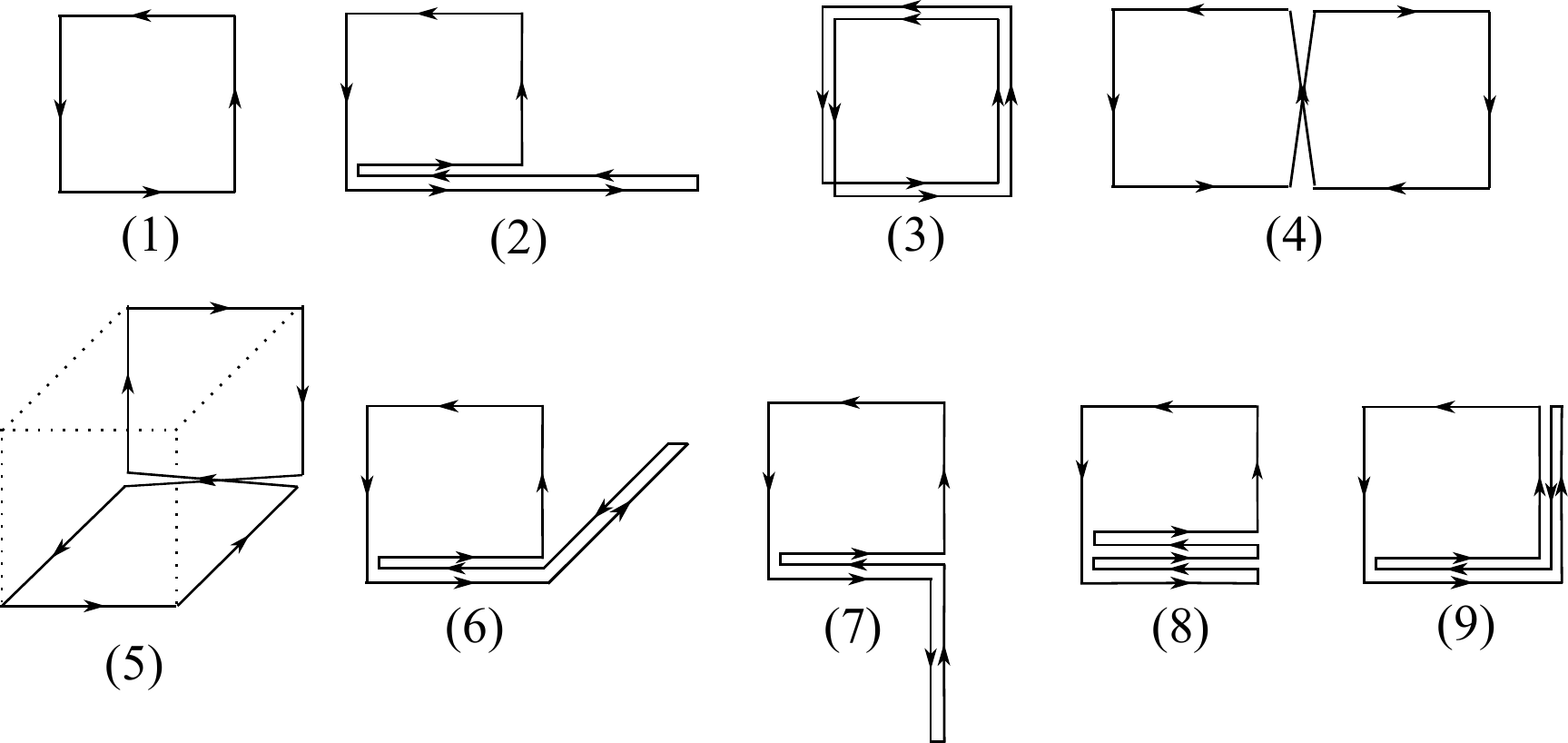}
	\caption{The Wilson Loops necessary for the plaquette loop equation of the bosonic \Nfour action}
	\label{fig:n4le_plaq}
\end{figure}

To test the loop equations we considered the loop equation associated with the plaquette 
and found that it is satisfied up to four digits which is within the numerical accuracy of the simulation. 
The test was done for U(2) and U(3) gauge groups and for $\lambda_{lat} = 0.8,1.0,1.2$ on
$8^4$ lattices.  The results are presented in table \ref{N4loops}.
\begin{table}[h]
	\centering
	\renewcommand{\arraystretch}{1.2}
	\begin{tabular}{@{}ccccccc@{}}
		\toprule
		&  \multicolumn{3}{c}{N=2}  & \multicolumn{3}{c}{N=3} \\
		$\lambda =$    & $0.8$ & $1.0$ & $1.2$ & $0.8$ & $1.0$ & $1.2$ \\
		\midrule
		$\cW_1$ & $0.01158$ &  $0.01451$  &   $0.01740$  &  $0.01151$ &  $0.01440$ & $0.01724$       \\
		$\cW_2$ & $0.00241$ &  $0.00377$  &   $0.00543$  &  $0.00241$ &  $0.00377$ & $0.00540$       \\
		$\cW_3$ & $0.00052$ &  $0.00082$  &   $0.00118$  &  $0.00048$ &  $0.00075$ & $0.00109$       \\
		$\cW_4$ & $0.00026$ &  $0.00041$  &   $0.00059$  &  $0.00026$ &  $0.00040$ & $0.00057$       \\
		$\cW_5$ & $0.00023$ &  $0.00036$  &   $0.00051$  &  $0.00023$ &  $0.00036$ & $0.00051$       \\
		$\cW_6$ & $0.00205$ &  $0.00321$  &   $0.00462$  &  $0.00206$ &  $0.00321$ & $0.00462$     \\  
		$\cW_7$ & $0.00218$ &  $0.00343$  &   $0.00494$  &  $0.00219$ &  $0.00341$ & $0.00489$       \\
		$\cW_8$ & $0.00525$ &  $0.00823$  &   $0.01184$  &  $0.00520$ &  $0.00812$ & $0.01164$       \\
		$\cW_9$ & $0.00330$ &  $0.00518$  &   $0.00745$  &  $0.00322$ &  $0.00503$ & $0.00723$       \\
		\midrule
		Eq.  & $-0.00004$ &  $-0.00004$  & $-0.00003$  &  $0.00002$ &  $-0.000003$ & $-0.00001$   \\
		\bottomrule
	\end{tabular}
	\caption{Wilson loop expectation values used to check the loop equation (\ref{eq:n4le_plaq})}
	\label{N4loops}
\end{table} 
The loop equation necessary for the plaquette with the pure bosonic action is given by eq. (\ref{eq:n4le_plaq}) and seen pictorial in fig. \ref{fig:n4le_plaq}
\beq\label{eq:n4le_plaq}
\frac{1}{2\lambda_{lat}}\left(-\cW_2 - \cW_3 - \cW_4 - 6\cW_5 +6\cW_6 +  \cW_7 + \cW_8 +  \cW_9\right) - \cW_1 = 0\ .
\eeq
This concludes our brief study of the \N{4} case. The next step, that we leave for future work, 
would be to investigate larger Wilson Loops with and without self-intersections and the generalization to the fermionic sector. 

\section{Conclusions} 
\label{conclusions}

The loop equation has been traditionally seen as a promising way to describe gauge theories in terms of gauge invariant quantities. In this paper we agree with this perspective but also point out that such equation has infinite solutions that have to be constrained by the condition that certain matrices are positive definite. At strong coupling such extra conditions are not necessary and therefore seem to have been largely ignored. On the other hand, in the physically relevant region of small coupling such conditions are crucial to obtain the correct solution. In fact they also give many constraints and properties that are actually independent of the action. In two dimensions this leads to a simple numerical procedure that reproduces the exact solution for any coupling.  This method can potentially be used for other two dimensional actions where the exact solution is not known but, more importantly, it can be extended to higher dimensions. The simple idea is that given two points in space and a set of open lines $\ell=1\ldots L$ between them, we can define an $L\times L$ matrix of closed loops where the entry $\ell\ell'$ is the expectation value of the closed loop obtained by going along path $\ell$ and returning along path $\ell'$. Such matrix has to be positive definite and can be considered as a reduced density matrix due to tracing over the color indices. Its entropy measures how much information is lost by taking the traces and gives a qualitative idea of the entropy of the system. The reason is that the entropy of the matrix $\rhoM$ vanishes in the ground state when all links are equal to the identity (up to gauge transformations) and is maximal at large coupling when the links fluctuate randomly. 

 Numerically we implement a procedure to solve the loop equation under the condition that the constraints are satisfied. It reproduces known results from Monte Carlo simulations but cannot be considered an improvement. It is possible that more computational resources could lead to better results especially 
 in the small coupling region relevant to the continuum limit. Also, it might be possible to improve the choice of the constraints, namely the matrix $\rhoM$, and/or use improved actions that approach the continuum limit faster.  
 It is also interesting to extend this method to other theories whose action can be formulated entirely in terms of closed loops. One such theory is \N{4} SYM, of particular importance since it plays a central role in the AdS/CFT correspondence. Initial steps in this direction suggest that the method can be applied but requires a better understanding of the fermionic loops. Another system to consider is three dimensional Yang-Mills where there are other approaches \cite{KN,3d}.

\section{Acknowledgments} \label{ack}

 This work was supported in part by the DOE through grant \protect{DE-SC0007884}. 
 We are very grateful for numerous discussions with P. Vieira, S. Catterall, D. Schaich on the matters of this work and/or lattice gauge theory in general. Also S.Catterall and D. Schaich graciously provided their lattice code allowing us to check the loop equations in the bosonic sector of \N{4} SYM.  We are also grateful to D. Minic and A. Jevicki for useful comments on a previous version of this work. 
 P.D.A. would like to thank the Wigner GPU Laboratory at the Wigner Research
 Center for Physics (Budapest, Hungary) for providing GPUs computer resources. He would also like to
 thank G. G. Barnaf\"oldi, M. F. Nagy-Egri, D. Ber\'enyi, and Z. Bajnok for helpful discussions.
  M.K. wants to thank the hospitality of the Perimeter Institute, Waterloo, CA and the SAIFR Institute (S\~ao Paulo, Brazil),  while part of this work was being done.

\section{Appendix: Semi-Definite programming (SDP)} \label{app}

Semi-definite programming \cite{sdp} is a type of optimization problem that has been the focus of a lot of attention recently in relation to problems in finance, engineering, and more recently has proved invaluable in the bootstrap program of conformal field theories (see \eg\ \cite{sdpaCFT}).

It can be stated very simply as:

Given $m$ real numbers $c_{i=1\ldots m}\in \mathbb{R}$ and $m$ real, symmetric $n\times n$ matrices $\mathbb{F}_{i=0\ldots m}\in \mathbb{R}^{n\times n}$, 
find $x_{i=1\ldots m}\in\mathbb{R}$  that minimize $\sum_{i=1}^m c_i x_i$ under the constraint that $\mathbb{X}=\sum_{i=1}^m x_i \mathbb{F}_i - \mathbb{F}_0$  is positive semi-definite ($\mathbb{X}\succeq 0$). 

The main observation in this field is that the space of semi-definite matrices is a convex cone in the space of all symmetric $n\times n $ matrices. Indeed, given two positive semi-definite matrices $\mathbb{X}_{1,2} \succeq 0$, namely
$y^t \mathbb{X}_{1,2} y\ge 0, \ \forall \ y\in\mathbb{R}^n$, then it is clear that $y^t (\alpha_1 \mathbb{X}_1 +\alpha_2 \mathbb{X}_2) y \ge 0$ for any real $\alpha_{1,2} \ge 0$. Thus, $\alpha_1 \mathbb{X}_1 + \alpha_2 \mathbb{X}_2 \succeq 0, \ \forall\ \alpha_{1,2}\ge0$ showing that positive semi-definite matrices form a convex cone. The condition that $\mathbb{X}$ belongs to the linear subspace generated by the $\mathbb{F}_{i=1\ldots m}$ shifted by $\mathbb{F}_0$ 
defines an intersection between this hyperplane and the semi-definite cone. This is a convex region over which we minimize a linear function. Therefore, the minimum is unique and should be located at the boundary of the domain, namely when the matrix $\mathbb{X}$ has at least one zero eigenvalue. The problem is then very similar to the more traditional problem of linear programming where one minimizes a linear function $\sum_{i=1}^m c_i x_i$ over the convex cone $y_{\ell=1\ldots n}\ge 0$ where the $y_\ell=\sum_{i=1}^m a_{\ell i} x_i$ for some given coefficients $\a_{\ell i}$.  

 There are many other problems that can be reduced to an SDP problem. For example, in section \ref{2D2} we need to solve 
\beqa
 &\mbox{Minimize}&\ \  S = -\frac{1}{\lambda} \cW_1 + \sum_{n=1}^{\infty} \frac{1}{n} \cW_n^2 \ ,\non \\
 &\mbox{such that}&  \rhoM^{(L)} = \frac{1}{L} \mathbb{T}[\cW_0,\ldots,\cW_L] \succeq 0\ ,
 \label{b1}
\eeqa
this is  problem of quadratic programming that can be reduced to an SDP problem by defining a new variable $t$ and imposing
\beq
  -\frac{1}{\lambda} \cW_1 + \sum_{n=1}^{\infty} \frac{1}{n} \cW_n^2 \le t \ ,
 \label{b2}
\eeq
or equivalently
\beq
 S_M = \left(\begin{array}{ccccc}
   t & (u-\frac{1}{2\lambda}) & \frac{\cW_2}{\sqrt{2}} & \frac{\cW_3}{\sqrt{3}} & \ldots \\
  (u-\frac{1}{2\lambda})  & 1 & 0 &0 & \ldots \\
    \frac{\cW_2}{\sqrt{2}} & 0 & 1 &0 &\ldots\\
     \frac{\cW_3}{\sqrt{3}} & 0 & 0 &1 &\ldots\\
     \vdots &  \vdots &  \vdots &  \vdots & \ddots
 \end{array}\right) \succeq 0\ .
 \label{b3}
\eeq
Thus the problem (\ref{b1}) is equivalent to
\beq
\mbox{Minimize}\ t, \mbox{such that} \ \ \rho^{L}\succeq 0, \ S_M\succeq 0 \ .
\label{b4}
\eeq
Once the problem has been casted as an SDP problem, there are several available packages that can be used to solve it. We found that for rapid development of small problems the matlab package cvx \cite{cvx} was very convenient and, for larger problems sdpa \cite{sdpa} was a good choice.

\end{document}